\documentclass[a4paper,fleqn]{cas-sc}

\usepackage[authoryear,longnamesfirst]{natbib}
\usepackage{graphicx}
\usepackage{booktabs}
\usepackage{multirow}
\usepackage{longtable, array}

\usepackage{changes}
\usepackage{pgfplots}
\pgfplotsset{compat=1.18}
\usepackage{tikz}
\usetikzlibrary{positioning,arrows.meta,fit,calc}
\usepackage{geometry}
\usepgfplotslibrary{statistics}
\usepackage{chngcntr}
\usepackage{placeins}   

\usepackage[utf8]{inputenc}
\usepackage[T1]{fontenc}
\usepackage{CJKutf8}

\usepackage{longtable}

\AtBeginDocument{%
  \counterwithout{figure}{section}%
  \counterwithout{figure}{subsection}%
  \counterwithout{figure}{subsubsection}%

  \counterwithout{table}{section}%
  \counterwithout{table}{subsection}%
  \counterwithout{table}{subsubsection}%

  \renewcommand{\thefigure}{\arabic{figure}}%
  \renewcommand{\thetable}{\arabic{table}}%
}

\def\tsc#1{\csdef{#1}{\textsc{\lowercase{#1}}\xspace}}
\tsc{WGM}
\tsc{QE}

\begin{document}
\let\WriteBookmarks\relax
\def\floatpagepagefraction{1}
\def\textpagefraction{.001}

\setlength{\tabcolsep}{5pt}

\shorttitle{From School AI Readiness to Student AI Literacy}

\shortauthors{Guan et al.}  

\title [mode = title]{From School AI Readiness to Student AI Literacy: A National Multilevel Mediation Analysis of Institutional Capacity and Teacher Capability}


%

\author[1]{Xiu Guan}
\author[1]{Mingmin Zheng}
\author[2]{Dragan Gašević}
\author[1]{Wenxin Guo}
\author[1]{Yingqun Liu}
\author[1]{Xibin Han}
\author[3]{Danijela Gasevic}
\author[4]{Ruiling Ma}
\author[1]{Qi Wu}
\author[1,2]{Lixiang Yan\corref{cor1}}

\cortext[cor1]{Corresponding author. Address: School of Education, Tsinghua University, Beijing, China.}

\affiliation[1]{organization={School of Education, Tsinghua University},
            city={Beijing},
            country={China}}

\affiliation[2]{organization={Faculty of Information Technology, Monash University},
            country={Australia}}

\affiliation[3]{organization={School of Public Health and Preventive Medicine, Monash University},
            country={Australia}}

\affiliation[4]{organization={Yunnan Chinese Language and Culture College, Yunnan Normal University},
            country={China}}         

\begin{abstract}
Artificial intelligence (AI) is increasingly embedded in vocational education systems, yet empirical evidence linking institutional AI readiness to student learning outcomes remains limited. This study develops and tests a 2-2-1 cross-level mediation framework examining how school-level AI readiness is associated with student AI literacy through aggregated teacher mechanisms. Using linked survey data from 1,007 vocational institutions, 156,125 teachers, and 2,379,546 students nationwide, multilevel models were estimated to assess direct, indirect, and contextual effects. Results indicate that overall school AI readiness is positively associated with student AI literacy after adjusting for institutional and regional characteristics. When examined independently, all readiness dimensions show positive associations, while simultaneous modelling suggests that readiness operates as an integrated organisational configuration. Cross-level mediation analyses reveal that aggregated teacher-perceived AI capability partially mediates the relationship between institutional readiness and student literacy, whereas general attitudinal acceptance measures do not demonstrate stable transmission effects. Robustness analyses further show that this readiness-capability-literacy pathway remains structurally stable across heterogeneous regional AI development contexts and under alternative modelling specifications. These findings reposition institutional AI readiness as a multilevel organisational condition linked to student AI literacy, identify collective teacher capability as its central transmission mechanism, and underscore the need to align infrastructural investment with sustained professional capacity development.

\end{abstract}



\begin{keywords}
AIED \sep AI literacy \sep  Institutional AI readiness \sep Vocational education \sep 21st century abilities 

\end{keywords}

\maketitle

\section{Introduction}

Artificial intelligence (AI) integration in education represents not merely a technological upgrade but a systemic organisational transformation that requires coordinated institutional capacity building. International frameworks consistently emphasise that AI’s educational potential depends on governance structures, data infrastructures, ethical safeguards, and sustained professional development rather than on tool availability alone \citep{OECD2026,OECD2023,UNESCO2023,holmes2022state}. In this context, AI literacy has emerged as a critical student-level outcome, particularly in vocational education where learners must apply AI tools within authentic occupational tasks \citep{unesco2024ai_students,wuttke2020vocational}. However, while policy discourse foregrounds institutional readiness as foundational, empirical evidence directly linking school-level AI capacity to measurable student competencies remains limited. The central unresolved question is not whether AI infrastructure exists, but how institutional readiness translates into tangible learning outcomes and through which organisational mechanisms this translation occurs.

Vocational education provides a particularly consequential setting for examining this institutional-instructional-student linkage because it sits at the intersection of schooling and rapidly evolving labour markets. The diffusion of AI across industrial production, logistics, healthcare, and service sectors has intensified pressure on vocational institutions to cultivate graduates who can critically evaluate, adapt, and responsibly apply AI systems in domain-specific workflows \citep{mikkonen2017guiding,NCVER2025employers,Billett2011}. Achieving this goal requires more than individual teacher enthusiasm or student exposure to tools; it demands systemic alignment among institutional strategy, governance, infrastructure, and pedagogical practice. Research from digital competence and technology-enhanced learning has long demonstrated that school-level structural conditions delimit the instructional opportunities available to students \citep{warschauer2010new,scherer2019technology}. Yet in AI-in-education scholarship, institutional factors are frequently treated as contextual background variables rather than as core explanatory constructs shaping student-level outcomes.

Existing research on AI in education remains fragmented across analytical levels, limiting theoretical integration and explanatory power. Student-focused studies typically conceptualise AI literacy as an individual competence or attitudinal construct \citep{long2020what}, whereas teacher-centred research largely examines adoption intentions, perceived usefulness, or acceptance dynamics \citep{celik2022promises,Ng2021,miao2024ai}. Parallel streams of descriptive work document variations in organisational AI readiness, including differences in infrastructure maturity, governance design, and professional development provision \citep{Karan07062025,angadi2026incorporation}. However, these strands rarely converge within a unified multilevel framework capable of tracing how macro-level institutional configurations influence meso-level instructional conditions and ultimately micro-level student competencies. Without modelling such cross-level pathways, institutional readiness remains conceptually suggestive but empirically under-specified.

Advancing the field requires an explicitly multilevel explanatory model that connects institutional AI readiness to student AI literacy through identifiable organisational transmission mechanisms. Organisational support theory suggests that institutional investments influence outcomes indirectly through professional intermediaries \citep{Rhoades2002}, while ecological systems theory posits that meso-level environments mediate the relationship between structural conditions and individual development \citep{Bronfenbrenner1994}. Applying these perspectives to AI integration implies that school-level readiness should shape aggregated teacher instructional capacity, which in turn structures students’ opportunities to develop AI literacy. Empirically testing this proposition demands large-scale, hierarchically structured data capable of estimating cross-level mediation rather than relying on single-level regression or small-sample case studies. Moreover, establishing the robustness of institutional effects requires examination across heterogeneous regional ecosystems. Much educational technology research has been criticised for contextual fragility, wherein findings derived from technologically advantaged environments fail to generalise to diverse or resource-constrained systems \citep{zawacki2019systematic}. This concern is especially salient in vocational education, where regional economic development, proximity to AI industries, and policy investment vary substantially \citep{oecd2019skills,unesco2022tvet}. Determining whether institutional readiness effects persist across divergent AI development contexts is therefore essential for informing equitable, system-level policy decisions.

Addressing these theoretical and empirical gaps, the present study develops and tests a 2-2-1 cross-level mediation framework linking school-level AI readiness, aggregated teacher instructional mechanisms, and students’ AI literacy in vocational education. Using a uniquely large linked dataset encompassing over two million students, more than one hundred thousand teachers, and over one thousand vocational institutions nationwide, the study models whether macroscopic disparities in institutional readiness are associated with collective teacher capability and whether this collective capacity mediates student literacy outcomes. By integrating multilevel modelling, mediation analysis, and contextual robustness testing, the study moves beyond descriptive readiness audits to provide mechanism-based evidence on how institutional AI capacity becomes educationally consequential. Guided by this framework, the study addresses three research questions:

\textbf{RQ1:} To what extent is school AI readiness associated with students’ AI literacy in vocational education, and which dimensions of institutional readiness exhibit consistent explanatory power?

\textbf{RQ2:} Through which aggregated teacher-level instructional mechanisms is school AI readiness associated with students’ AI literacy, and do these mechanisms constitute identifiable transmission pathways?

\textbf{RQ3:} To what extent do regional AI development conditions moderate the cross-level associations between school AI readiness, teacher mechanisms, and student AI literacy, and are the observed pathways robust across alternative contextual specifications?

\section{Theoretical Framework and Literature Review}

\subsection{A Multilevel Ecological Framework for AI Integration}

AI integration in vocational education must be understood as a multilevel ecological process in which institutional, professional, and classroom systems interact to shape student learning outcomes. Drawing on ecological systems theory \citep{Bronfenbrenner1979,Bronfenbrenner2005}, educational environments are conceptualised as nested structures comprising macro-level policy and industry contexts, meso-level institutional organisations, and micro-level classroom interactions. Within this hierarchy, macro-level forces such as regional digitalisation, labour-market transformation, and national AI strategies create structural conditions for institutional action. Meso-level schools translate these external pressures into governance arrangements, resource allocation, and professional development systems. Micro-level instructional processes, where teachers, students, and AI tools interact, constitute the proximal mechanisms through which competencies are formed. This nested architecture implies that AI-related educational phenomena are inherently cross-level and therefore require analytical models capable of tracing institutional influences through organisational mediators to individual student outcomes.

Recent AI-in-education syntheses increasingly argue that educational impact depends less on technological availability and more on organisational capacity and pedagogical alignment \citep{Bond2024,Wu2024}. In vocational education specifically, AI tools must be embedded within practice-oriented curricula and authentic occupational tasks \citep{Billett2011}. This requirement intensifies the importance of institutional coordination because misalignment between infrastructure and instructional design may render AI resources pedagogically inert. Schools thus function as meso-level configurators of instructional opportunity structures, determining whether AI integration manifests as isolated tool usage or as systematically scaffolded competency development \citep{Spillane2002}. Teachers, in turn, act as the critical mediators translating institutional strategy into classroom enactment \citep{Goddard2026,Emam2026EvaluatingIS,miao2024ai}. Conceptually, this perspective positions AI readiness not as a background condition but as an organisational force that shapes the conditions under which teaching and learning occur.

\subsection{Institutional AI Readiness as Organisational Capacity}
\label{subsec:school_readiness}

Institutional AI readiness can be conceptualised as a collective organisational capacity to coordinate strategy, infrastructure, governance, and professional development for sustained innovation. Drawing on organisational readiness for change theory \citep{weiner2009organizational,weiner2020theory} and the Technology-Organisation-Environment (TOE) framework \citep{Tornatzky1990}, readiness reflects not only technical preparedness but also shared commitment, structural alignment, and implementation efficacy. Within vocational education, these elements determine whether AI integration coheres with applied assessment demands, regulatory expectations, and industry standards. Thus, readiness is not merely a technical condition but a systemic organisational attribute shaping the feasibility and coherence of AI-enabled pedagogy.

This study operationalises school-level AI readiness as a multidimensional configuration encompassing strategic and policy readiness, organisational environment, process readiness, technological readiness, data readiness, and ethical governance \citep{ali2024ai,johnk2021ready,karan2025understanding}. These components capture both structural and motivational prerequisites for AI-supported transformation. Importantly, ecological theory suggests that such dimensions co-occur and function synergistically rather than independently \citep{Bronfenbrenner1994}. For example, technical infrastructure without governance clarity may produce risk, while ethical policy without data capacity may impede implementation. Consequently, overall readiness reflects systemic coherence, whereas individual dimensions may exhibit varying proximity to classroom instruction and thus differential associations with student outcomes \citep{filderman2022data,konstantinidou2022teaching}. Grounded in this configurational logic, we hypothesise:

\textbf{H1a:} Overall school-level AI readiness is positively associated with students' AI literacy.

\textbf{H1b:} The associations between individual school-level AI readiness dimensions and students' AI literacy vary in magnitude.

\subsection{Aggregated Teacher Capability as the Organisational Transmission Mechanism}

Institutional conditions influence student outcomes primarily through professional intermediaries who enact instructional practices within classrooms. Organisational support theory posits that institutional investments shape employee perceptions of capability and support, which in turn affect behavioural engagement and performance \citep{Rhoades2002}. Applied to AI integration, school-level readiness should influence teachers’ collective perceptions of their instructional capacity to design and implement AI-supported pedagogy. Without such professional translation, institutional resources risk remaining structurally present but pedagogically inactive.

Technology adoption frameworks such as the Unified Theory of Acceptance and Use of Technology (UTAUT) provide insight into how expectancy beliefs and facilitating conditions shape technology use \citep{venkatesh2003user,williams2015unified}. However, vocational AI integration extends beyond individual adoption decisions and requires coordinated, curriculum-embedded instructional enactment. Recent studies suggest that domain-specific pedagogical competence, rather than general attitudinal acceptance, is central to effective AI implementation \citep{taheri2025factors,Cui06112025}. Evidence from large-scale ICT research further demonstrates that aggregated teacher contexts condition students’ technological competencies \citep{kastorff2024teachers,lohr2024digital}. When teachers collectively perceive themselves as capable of integrating AI into industry-aligned tasks, they are more likely to orchestrate explicit literacy development opportunities \citep{zhang2024effectiveness}. Accordingly, this study conceptualises teacher mechanisms as aggregated professional conditions at the school level, capturing both perceived AI capability and experiential evaluations of AI integration (e.g., expectations, effort, institutional support). These aggregated constructs represent organisational climate rather than individual idiosyncrasy, thereby aligning conceptually with the meso-level role posited in the ecological framework. We therefore hypothesise:

\textbf{H2:} Aggregated teacher-level mechanisms mediate the positive association between school-level AI readiness and students' AI literacy.

\subsection{Student AI Literacy and Contextual Robustness of Institutional Effects}

Student AI literacy represents a multidimensional competency encompassing foundational knowledge, applied tool use, critical evaluation, and ethical reasoning \citep{ng2021ailiteracy,UNESCO2023}. In vocational education, this competency is expressed through authentic occupational problem-solving in AI-enriched environments \citep{lin2025artificial,metreveli2025will}. Research consistently indicates that such competencies emerge not from device exposure alone but from structured, pedagogically guided integration \citep{scherer2019technology}. Consequently, if institutional readiness meaningfully shapes classroom instruction through teacher capability, it should manifest in measurable differences in student AI literacy. However, the strength and stability of institutional effects must be evaluated across heterogeneous macroeconomic environments. Vocational systems differ substantially in regional AI industry maturity, digital infrastructure, and socioeconomic development \citep{oecd2019pisa2018v1,oecd2019skills}. Educational technology literature cautions against contextual fragility, whereby findings derived from advantaged environments fail to generalise \citep{zawacki2019systematic}. To establish scalability, it is therefore necessary to test whether the readiness-capability-literacy pathway remains structurally stable across regions with differing AI development conditions. Accordingly, we hypothesise:

\textbf{H3:} The cross-level mediated association between school-level AI readiness and students' AI literacy demonstrates structural robustness and does not vary substantively across regional AI development contexts.

\section{Method}

This study analysed linked, multi-respondent data from a large-scale national survey of vocational education institutions in China. The design was cross-sectional; accordingly, all findings are interpreted as associations rather than causal effects. The analytic structure was hierarchical: students (Level~1) were nested within schools (Level~2). Teacher responses were aggregated to the school level to represent shared institutional instructional and organisational conditions experienced by students within the same school.

\subsection{Sample and Procedures}

Data were collected using three parallel online questionnaires administered to schools, teachers, and students. All procedures were reviewed and approved by the ethics committee of [Blinded] University. Participation was voluntary, and informed consent was obtained electronically prior to survey completion. Responses were anonymised at the school level using unique identifiers, and no personally identifiable information was collected or stored. The survey was conducted between March~25 and April~10, 2025, and covered all 31 provincial-level administrative regions in China. The sampling frame was constructed from national vocational education statistics published by the Ministry of Education of China \citep{MOE2023statistics} and encompassed the nationwide population of vocational institutions.

\paragraph{Sampling design.}
A stratified proportional sampling strategy was implemented at the provincial level to enhance representativeness and regional balance, following standard survey methodology principles \citep{Yuan2013methods}. The target number of sampled institutions in each province was set proportional to the provincial population of vocational institutions (including both secondary vocational and higher vocational schools). To ensure adequate coverage for multilevel and regional robustness analyses, predefined minimum sampling thresholds were also established for each province and institutional type. Within each province, schools were randomly invited through provincial education authorities and vocational education networks. Although a small number of provinces fell slightly below predefined minimum thresholds due to non-response or administrative constraints, the achieved sample distribution remained broadly aligned with the intended design. Detailed provincial sampling targets and achieved sample sizes by institutional type are reported in Appendix Table~\ref{tab:prov_sampling}.

\begin{figure}
    \centering
    \includegraphics[width=1\linewidth]{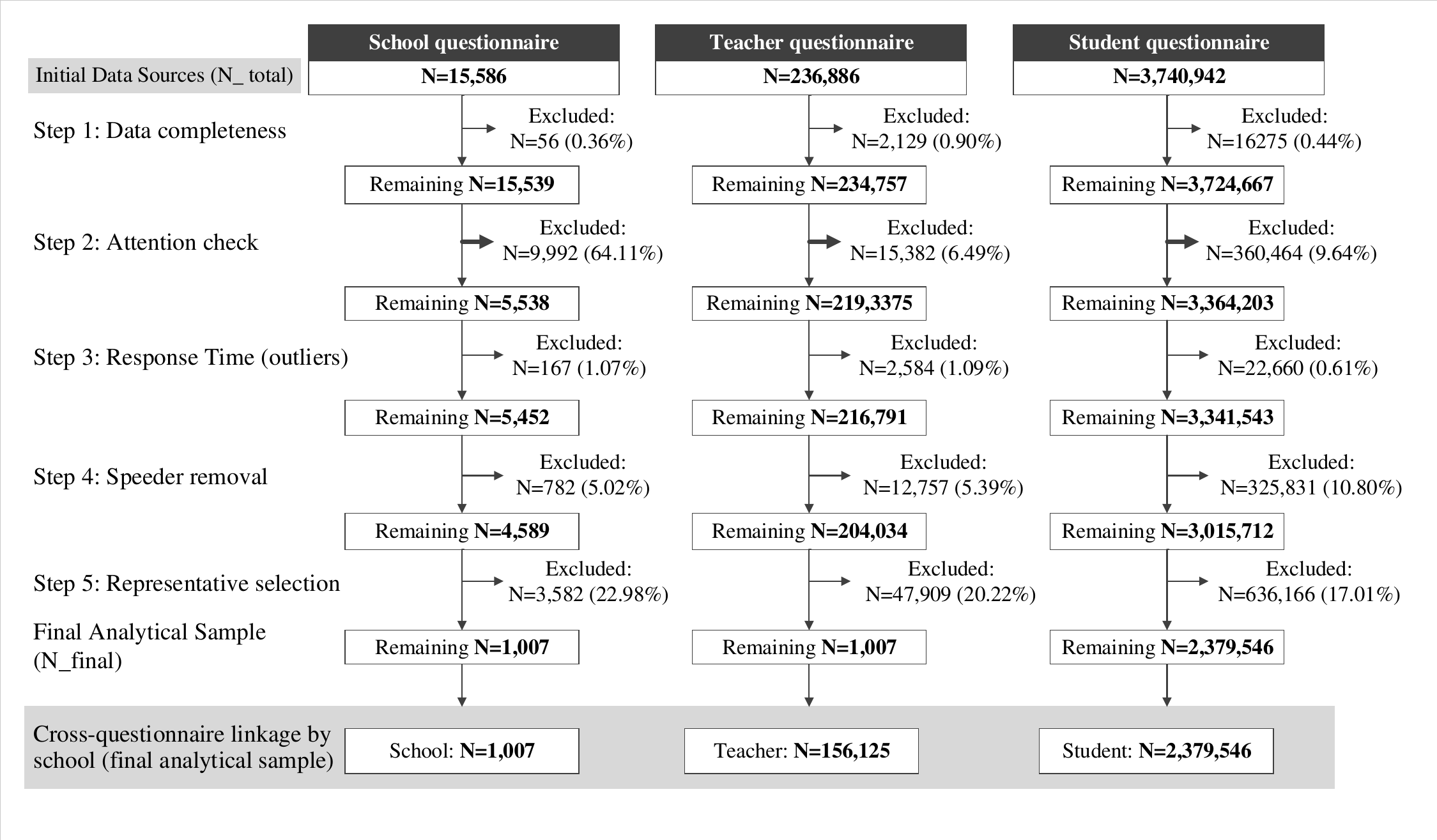}
    \caption{Data screening, linkage, and construction of the final analytical sample}
    \label{fig:dataflow}
\end{figure}

\paragraph{Data linkage and quality screening.}
As summarised in Figure~\ref{fig:dataflow}, data quality control was conducted separately for the school, teacher, and student questionnaires and then verified after cross-questionnaire linkage using unique school identifiers. Screening included: (a) completeness checks; (b) attention checks; (c) response-time outlier detection; and (d) identification of speeders. For the school questionnaire, additional procedures addressed duplicate institutional submissions. When duplicates were detected, responses were prioritised based on respondents’ organisational roles, with preference given to school leaders or administrators directly responsible for AI-related planning and implementation (e.g., principals, vice principals in charge of digital transformation, or heads of information technology units). If multiple submissions were from comparable administrative roles, the most complete response was retained. 

Attention checks were used as an initial reliability filter to identify inattentive or mechanically completed responses. The questionnaires included instructed-response and internal consistency check items (e.g., explicitly requesting respondents to select a specified option). Responses failing at least one predefined attention criterion were excluded prior to subsequent time-based screening. Completion times were trimmed at the 0.5th and 99.5th percentiles to remove extreme anomalies, then log-transformed to address positive skew common in response-time data. Cases exceeding $\pm 3$ standard deviations on the log-transformed scale were excluded as statistical outliers \citep{Berger2021}. Behavioural speeders were also removed. Following established web survey methodology \citep{malhotra2008completion,Zhang_Conrad_2014,greszki2015exploring}, respondents with total completion time less than one-third of the median completion time were classified as speeders and excluded.

To ensure adequate provincial coverage and balanced representation of secondary and higher vocational institutions, a limited recovery procedure was applied for a small number of schools in underrepresented regions, subject to strict quality criteria and cross-questionnaire consistency checks. After automated screening and manual verification, four schools met recovery criteria, yielding a final sample of 1,007 institutions. The full screening and linkage pipeline, including sample attrition at each stage, is summarised in Figure~\ref{fig:dataflow}. After linkage and quality screening, the final analytic dataset comprised $N_{schools}=1{,}007$ vocational institutions, $N_{teachers}=156{,}125$ teachers, and $N_{students}=2{,}379{,}546$ students.

\paragraph{Institutional policy status.}
Among the 1,007 schools, 283 institutions were selected under the National ``Double High'' Initiative, 85 under the Provincial ``Double High'' Initiative or other provincial excellence programs, 178 under the National ``Double Excellence'' Initiative, and 291 under the Provincial ``Double Excellence'' Initiative or related provincial programs. The remaining 370 institutions were classified as other vocational institutions. This distribution reflects substantial heterogeneity in institutional development status and policy engagement, supporting stratified analyses across different tiers of vocational institutions.

\paragraph{Teacher sample characteristics.}
The teacher sample included 56,111 male teachers and 100,014 female teachers. According to the 2020 National Education Development Statistics released by the Ministry of Education of China \citep{MOE2021NationalEduStatistics}, women constitute a substantial proportion of the vocational education workforce nationally. Recent sector-level survey reports similarly indicate a relatively high proportion of female teachers in vocational education settings \citep{EOL2023VocTeacherReport}. At the same time, survey methodology research suggests that women may exhibit higher response rates in mail and web-based surveys \citep{becker2022gender}. Therefore, while the observed gender distribution is consistent with national patterns, potential differential response rates should be considered when interpreting sample composition. In terms of teaching experience, 13,081 teachers had less than one year of experience, 24,694 had one to three years, 35,298 had four to ten years, and 83,052 had more than ten years. This distribution includes early-career, mid-career, and senior teachers, supporting the stability of aggregated teacher indicators at the school level.

\begin{figure}
    \centering
    \includegraphics[width=1\linewidth]{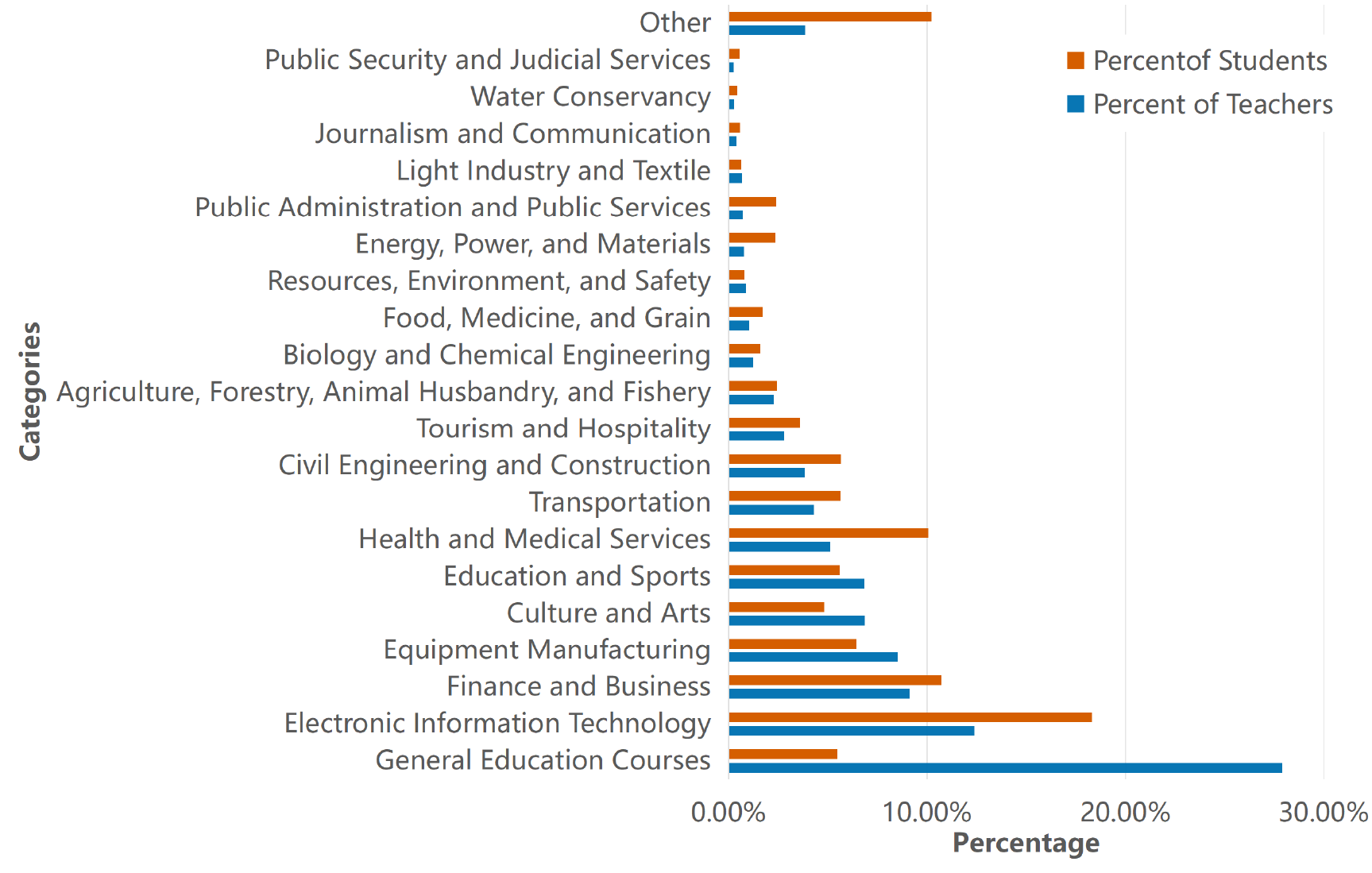}
    \caption{Distribution of student and teacher samples across vocational program categories}
    \label{fig:datacourse}
\end{figure}

Teachers were classified by the vocational program category of the courses they primarily taught (Figure~\ref{fig:datacourse}). The sample covered a broad range of instructional domains, including general education, technology-intensive programs, manufacturing and engineering, business-related disciplines, health and service-oriented tracks, and public service-related programs. The distribution was compared with official national statistics on vocational teacher specialisation \citep{MOE2025VocationalTeacherStats}. The sample includes all principal professional categories observed nationally and shows a directionally consistent pattern, while allowing for proportional variation across specific fields.

\paragraph{Student sample characteristics.}
The student sample included 1,142,270 male students and 1,237,276 female students. Students were drawn from a wide range of vocational program domains, with strong representation from information technology-related fields, manufacturing and engineering, health and medical services, business and management, and transportation, construction, and education-oriented tracks (Figure~\ref{fig:datacourse}). Students from agriculture, energy and materials, public services, tourism, and other applied fields were also represented. This distribution reflects the multi-sectoral structure of China’s vocational education system and supports generalisability across heterogeneous occupational domains. The program distribution broadly aligns with national enrolment patterns based on official statistics \citep{MOE2024VocationalStudentStats}, with proportional differences described descriptively.

\subsection{Measures}

Consistent with the theoretical framework (Section~2), school-level AI readiness was operationalised as a contextual organisational condition (Level~2), teacher-level mechanisms as aggregated mediating processes (Level~2), and student AI literacy as the learning outcome (Level~1). The analytical framework was specified as a 2-2-1 cross-level mediation model, in which the predictor and mediators are at Level~2 and the outcome is at Level~1 \citep{preacher2010general,zhang2009testing}. In this specification, school AI readiness functions as a contextual predictor, aggregated teacher mechanisms represent school-level mediators, and student AI literacy is modelled at the individual level. Given the cross-sectional design, the measures support inference about associations rather than causal effects.

\subsubsection{Level~2 (School): AI readiness}

School AI readiness was measured using the school questionnaire. Scale development was grounded in organisational readiness for change theory \citep{weiner2009organizational,weiner2020theory} and informed by AI and digital readiness frameworks developed for public-sector and educational contexts \citep{ali2024ai,educause2023ai}. AI readiness was operationalised as six dimensions: (1) strategic and policy readiness, (2) organisational environment, (3) process readiness, (4) technological readiness, (5) data readiness, and (6) ethical governance. Higher scores indicate higher readiness. Dimension scores were aggregated to form an overall school AI readiness composite index representing the joint configuration of these institutional capacities.

Because multiple school-level constructs were collected via self-report within a single survey wave, potential common method bias was assessed. Harman’s single-factor test (unrotated exploratory factor analysis of all self-report items) indicated that the first unrotated factor explained 45.3\% of the total variance, which did not exceed the commonly referenced 50\% threshold. This pattern suggests that variance was not overwhelmingly attributable to a single common method factor. Full school questionnaire items are provided in Appendix Table~\ref{tab:school_questionnaire_1}--\ref{tab:school_questionnaire_9}. Internal consistency and convergent validity were evaluated using Cronbach’s $\alpha$, composite reliability (CR), and average variance extracted (AVE). Cronbach’s $\alpha$ ranged from 0.839 to 0.937 and CR ranged from 0.855 to 0.937, exceeding the recommended 0.70 threshold \citep{Hair2022PLSSEM}. AVE values exceeded the recommended 0.50 criterion, supporting convergent validity.

Confirmatory factor analyses were conducted to compare a single-factor model with the hypothesised six-factor measurement model using the WLSMV estimator. The six-factor model fit the data substantially better than the single-factor alternative ($CFI = 0.994$ vs.\ 0.978; $RMSEA = 0.106$ vs.\ 0.200), and a chi-square difference test supported the superiority of the hypothesised structure ($\Delta \chi^2 = 2833.83,\ p < 0.001$). All standardised factor loadings were statistically significant ($p < 0.001$) and ranged from 0.442 to 0.929. Although one loading was below the conservative 0.50 benchmark, all exceeded the 0.40 threshold commonly used for practical significance in applied measurement research \citep{Hair2011}. Taken together, these results support the intended multidimensional measurement structure and indicate that common method bias is unlikely to pose a serious threat to validity in the present analyses.

\subsubsection{Level~2 (Aggregated): teacher mechanisms}

Teacher mechanisms were measured using the teacher questionnaire and conceptualised as school-shared instructional and organisational processes through which institutional AI conditions are reflected in everyday teaching practice. The full list of teacher questionnaire items is provided in Appendix Tables~\ref{tab:teacher_questionnaire_1}--\ref{tab:teacher_questionnaire_4}. The measures capture teachers’ AI-related perceptions, usage tendencies, and perceived organisational and instructional support, reflecting how teachers experience, interpret, and respond to AI-related initiatives within their institutions.

Item development operationalised the two mediating dimensions defined in Section~2: (a) perceived AI capability and (b) experiential and evaluative orientations toward AI-supported teaching. To ensure conceptual alignment and measurement validity, item construction was informed by established technology adoption frameworks, particularly the Technology Acceptance Model (TAM) and the Unified Theory of Acceptance and Use of Technology (UTAUT). These frameworks provide validated structures to represent performance expectancy, effort appraisal, social influence, and facilitating conditions, which map onto the expectancy beliefs and enabling conditions articulated in the background section. Selected elements from the Artificial Intelligence--Driven Decision Aids (AIDUA) framework were incorporated to reflect AI-specific instructional decision contexts. In the present study, these frameworks served as measurement scaffolds for the theoretically defined mediators rather than independent theoretical foundations. The resulting factor structure and descriptive statistics are summarised in Table~\ref{tab:utaut_factors}. Internal consistency reliability was satisfactory (Cronbach’s $\alpha$ = 0.77--0.96; $\omega_{\text{total}}$ = 0.79--0.98).

\begin{table}[htbp]
\centering
\caption{UTAUT-Informed Factor Structure of AI-Related Teacher AI Acceptance}
\label{tab:utaut_factors}
\small
\setlength{\tabcolsep}{3pt}
\renewcommand{\arraystretch}{1.12}

\resizebox{\linewidth}{!}{%
\begin{tabular}{@{}p{3.5cm} p{5cm} c p{2.3cm} c c@{}}
\toprule
\textbf{UTAUT Construct} & \textbf{Factor} & $\boldsymbol{\alpha}$ & \textbf{Item No(s).} & \textbf{Mean} & \textbf{SD} \\
\midrule
\multirow{2}{*}{Performance Expectancy}
& Perceived Positive Effects of AI & 0.963 & T1.1 (8 items) & 3.898 & 0.163 \\
& Perceived Negative Effects of AI & 0.924 & T1.2 (7 items) & 3.545 & 0.133 \\
\midrule
\multirow{2}{*}{Effort Expectancy}
& Perceived Difficulty of Using AI & 0.772 & T1.3 (3 items) & 3.708 & 0.143 \\
& Satisfaction with AI Use in Teaching & 0.904 & T1.4 (4 items) & 3.089 & 0.153 \\
\midrule
Social Influence
& Perceived Social Support for AI-Enabled Teaching & 0.836 & T1.5 (2 items) & 3.954 & 0.196 \\
\midrule
Performance Expectancy
& Perceived AI Capability & 0.816 & T1.6 (4 items) & 3.622 & 0.147 \\
\midrule
Facilitating Conditions
& Perceived Supportive Conditions for AI Use & 0.850 & T1.7 (3 items) & 3.520 & 0.200 \\
\bottomrule
\end{tabular}%
}
\end{table}

Although these variables were collected from individual teachers, they were aggregated to the school level to represent shared instructional and organisational conditions within each institution. This aggregation is consistent with the educational AI ecosystem framework (Section~2), which conceptualises schools as meso-level organisational systems shaping common instructional environments, and with organisational support theory, which treats perceptions of support as climate-like properties that can be meaningfully characterised at the organisational level \citep{Rhoades2002}. Aggregation is widely used in multilevel educational and organisational research when constructs are theorised to represent collective conditions \citep{Bliese2000,Raudenbush2002,Goddard2026}. The appropriateness of aggregation was evaluated using intraclass correlation coefficients and within-group agreement indices. ICC(1) estimates the proportion of variance attributable to between-school differences, and ICC(2) reflects the reliability of school-level mean scores. Across teacher constructs, ICC(1) values ranged from 0.008 to 0.044, indicating modest but non-trivial between-school variability. ICC(2) values ranged from 0.563 to 0.878, indicating acceptable to strong reliability for aggregated school means. Within-group agreement ($r_{wg(j)}$) ranged from 0.603 to 0.967, supporting sufficient within-school consensus to justify aggregation for multilevel analyses \citep{Bliese2000}.

\subsubsection{Level~1 (Student): AI literacy}

Student AI literacy was measured using the student questionnaire and served as the primary Level~1 outcome. The measure was adapted from the Generative AI Literacy Assessment Test (GLAT) \citep{JIN2025100436} and contextualised for vocational education settings \citep{ng2021ailiteracy}. The assessment captured students’ competence in understanding, applying, evaluating, and ethically engaging with generative AI in learning contexts. The instrument comprised ten dichotomously scored items (1 = correct, 0 = incorrect), reported in Appendix Tables~\ref{tab:student_questionnaire_1}--\ref{tab:student_questionnaire_2}. Internal consistency was estimated using the Kuder--Richardson Formula 20 (KR-20), which is appropriate for binary items and equivalent to Cronbach’s $\alpha$ under dichotomous conditions. The KR-20 coefficient was 0.63. Although modest, this level of reliability is commonly considered acceptable for short, population-based educational assessments, particularly when the construct is multidimensional and items are intentionally heterogeneous in content \citep{Nunnally1994,DeVellis2017}.

Item quality was evaluated using classical test theory indices. Item difficulty (proportion correct) ranged from 0.082 to 0.667, indicating coverage from very difficult to moderately difficult items. Item discrimination (point-biserial correlations) ranged from 0.157 to 0.408. Most items showed acceptable to strong discrimination ($r_{pb} > 0.20$), with three items exceeding 0.35. One item (Item 2) had lower discrimination ($r_{pb} = 0.157$) and high difficulty (p = 0.082). Given the extremely large sample size (>$1.8$ million students) and the conceptual importance of the item content, it was retained. Total scores ranged from 0 to 10 (M = 3.008, SD = 1.240), with skewness = 0.077. Fewer than 15\% of students achieved the maximum score, indicating limited ceiling effects. The full score distribution is presented in Panel A of Figure~\ref{fig:school_readiness_descriptives}.

\subsection{Data Analysis}

\begin{figure}
    \centering
    \includegraphics[width=1\linewidth]{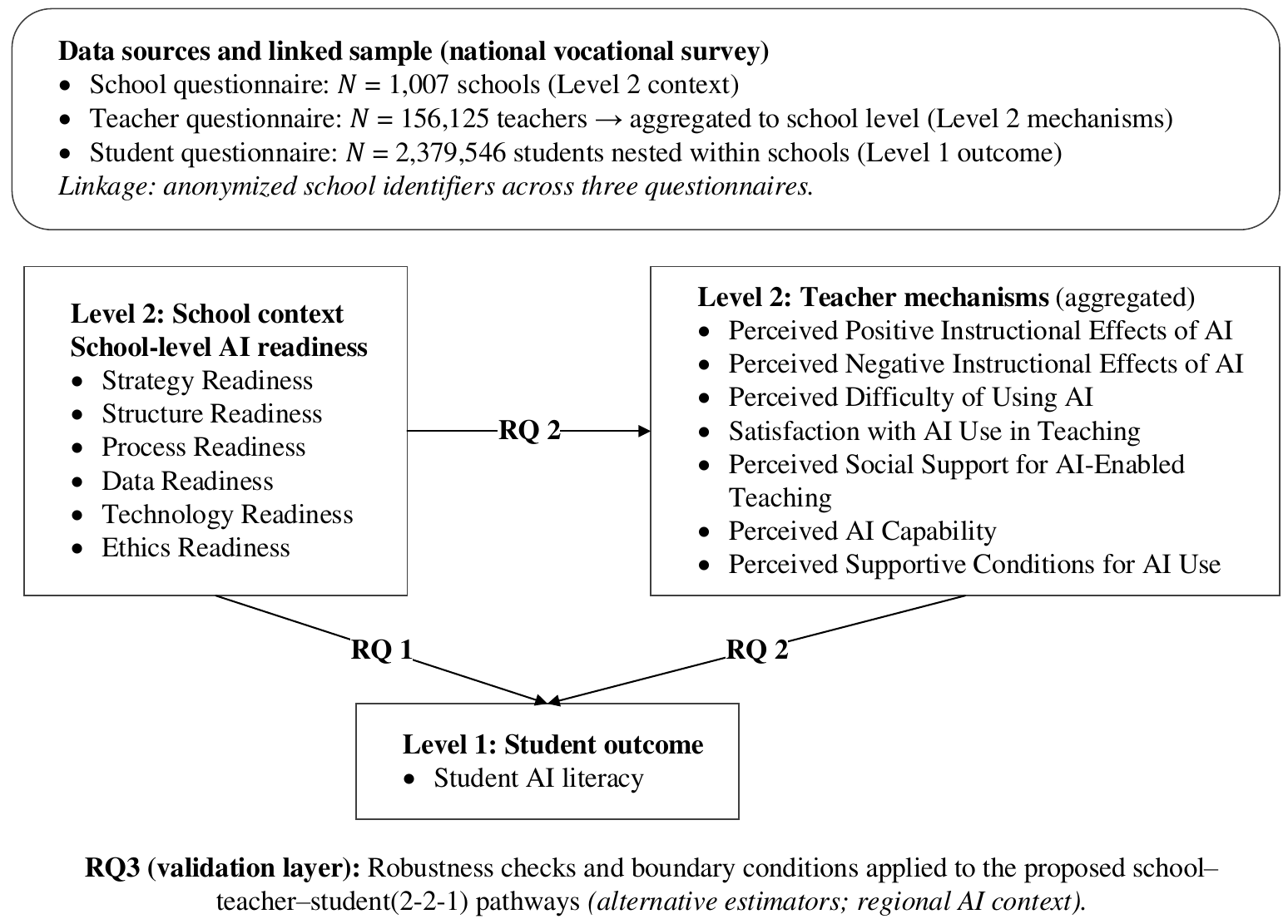}
    \caption{Cross-Level Mediation Framework Linking School AI Readiness, Aggregated Teacher Mechanisms, and Student AI Literacy}
    \label{fig:RQ123}
\end{figure}

All analyses were designed to align with the multilevel structure of the data and the hypothesised cross-level relationships among school AI readiness, aggregated teacher mechanisms, and student AI literacy. Given the nesting of students within schools, multilevel linear mixed-effects models with random intercepts at the school level were estimated to account for within-school dependency while enabling formal testing of cross-level direct and indirect effects. The conceptual multilevel framework is illustrated in Figure~\ref{fig:RQ123}. All continuous predictors were standardised prior to estimation to facilitate coefficient comparability. Fixed effects are reported as standardised regression coefficients ($\beta$), standard errors (SE), Satterthwaite-adjusted degrees of freedom, corresponding $t$ statistics ($t(df)$), and $p$ values. Indirect effects were evaluated using the product-of-coefficients approach. Because indirect effects typically exhibit non-normal sampling distributions, both the Delta method and Monte Carlo confidence intervals (20,000 draws) were used to assess statistical significance and robustness.

Preliminary analyses established the baseline variance structure and organisational-level associations prior to hypothesis testing. Descriptive statistics were computed at both student and school levels, and Spearman rank-order correlations were estimated among school-level AI readiness and aggregated teacher mechanisms ($N = 1{,}007$ schools). An unconditional (null) multilevel model was estimated to compute the intraclass correlation coefficient (ICC), quantifying the proportion of variance in student AI literacy attributable to between-school differences before inclusion of Level-2 predictors.

Hypothesis 1 was tested using covariate-adjusted multilevel models examining the association between school AI readiness and student AI literacy. All models included three structural school-level covariates: economic-administrative region (Eastern, Central, Western, Northeastern China), policy status (national/provincial key institution vs.\ non-designated), and institutional type (secondary vs.\ higher vocational). Separate models were first estimated for overall AI readiness and for each of the six readiness dimensions to assess total associations net of structural covariates. A simultaneous model including all six readiness dimensions was then estimated to evaluate their relative associations under shared variance conditions. Given conceptual overlap among readiness components, generalised variance inflation factors (GVIF) were computed to assess multicollinearity among Level-2 predictors.

Hypothesis 2 was examined using a 2-2-1 cross-level mediation framework in which overall school AI readiness was specified as the Level-2 predictor, aggregated teacher mechanisms as Level-2 mediators, and student AI literacy as the Level-1 outcome. The $a$ paths estimated associations between school AI readiness and aggregated teacher mechanisms at the school level. The $b$ paths estimated cross-level associations between aggregated teacher mechanisms and student AI literacy while controlling for school AI readiness and structural covariates. The direct effect ($c'$) represented the remaining association between school AI readiness and student AI literacy after accounting for mediators. Given the extremely large student-level sample size, $b$-path models were estimated using a stratified random subsample of up to 300 students per school to ensure computational feasibility while preserving the hierarchical structure. Indirect effects ($a \times b$) were evaluated using the Delta method and validated using Monte Carlo simulations (20,000 draws).

Hypothesis 3 was evaluated through complementary robustness and moderation analyses assessing contextual boundary conditions. First, an aggregated 2--2--2 mediation model was estimated at the school level, modelling school AI readiness, aggregated teacher mechanisms, and aggregated student AI literacy entirely at Level~2 to provide a conservative robustness benchmark. Second, moderation analyses tested whether regional AI development, operationalised using the provincial AI industry competitiveness index \citep{ccid2024ai}, altered the readiness--literacy association. Both categorical (high vs.\ low AI development) and continuous specifications were estimated using cross-level interaction terms, with simple slope analyses conducted where appropriate. Third, province-level fixed effects and cross-level interaction terms were incorporated to assess whether the readiness--literacy relationship varied systematically across provincial contexts. All moderation models adjusted for the same structural school-level covariates to ensure effects were estimated independently of baseline institutional heterogeneity.

\section{Results}

Given the multilevel structure of the data and the use of school-level predictors to explain student-level outcomes, the standardised coefficients reported below are expected to be modest in magnitude, which is typical in large-scale educational research \citep{kraft2020interpreting,tobler2024context,rose2001sick}.

\subsection{Descriptive statistics and preliminary associations}
\label{sec:results-des-stats}

\begin{figure}
    \centering
    \includegraphics[width=\linewidth]{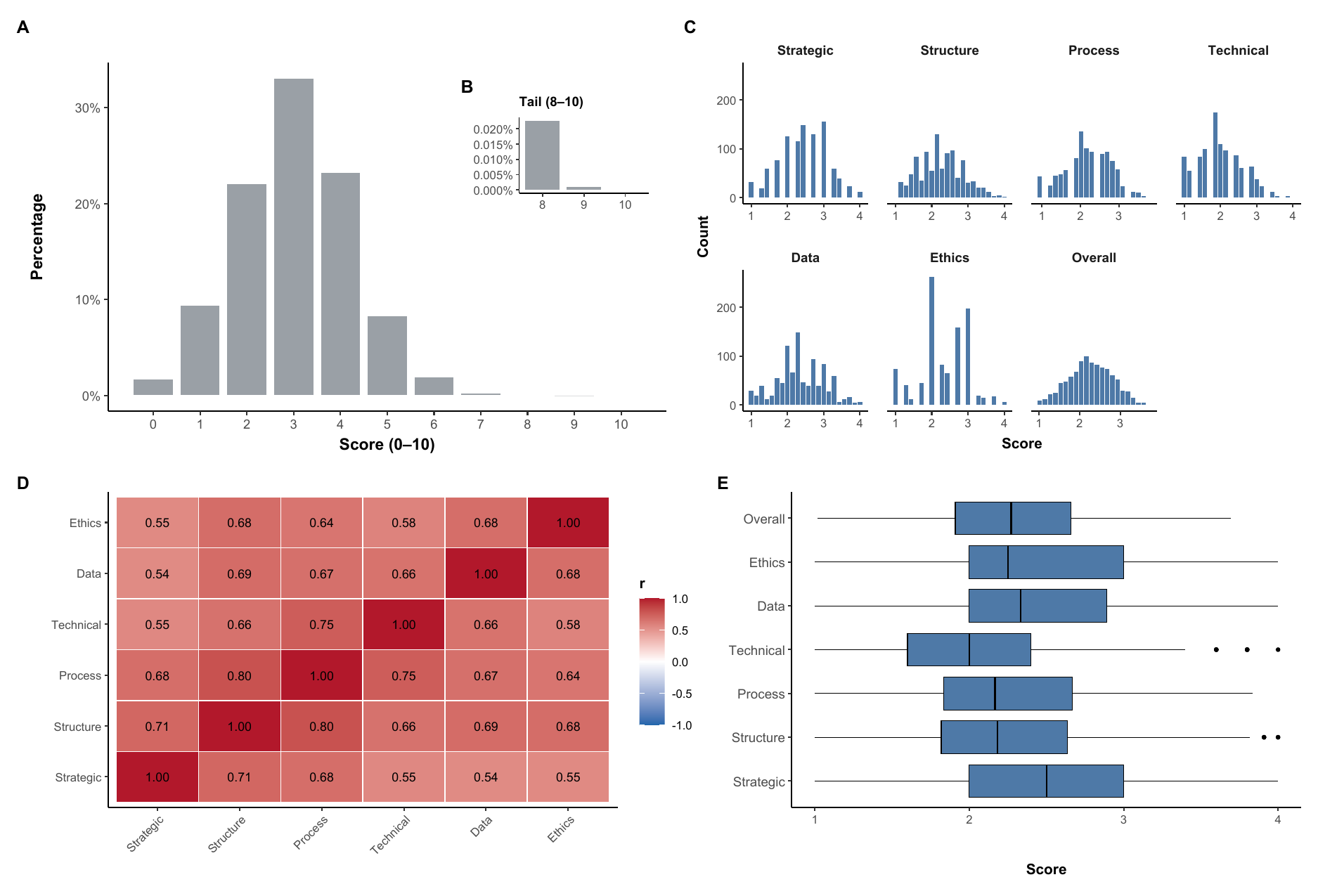}
    \caption{\textbf{Distributional characteristics and interrelationships of students' AI literacy and school-level AI readiness indicators.}}
    \label{fig:school_readiness_descriptives}
    \begin{flushleft}
    \footnotesize
    \textit{Note.} 
    \textbf{Panel A} shows the distribution of overall students' AI literacy scores (0--10 scale). 
    \textbf{Panel B} presents the magnified tail distribution of high AI literacy scores (8--10). 
    \textbf{Panel C} displays the score distributions of six readiness dimensions and the overall readiness index across schools. 
    \textbf{Panel D} illustrates the Pearson correlation matrix among readiness dimensions. 
    \textbf{Panel E} presents boxplots summarizing the distributional characteristics (median, interquartile range, and outliers) of overall readiness and six readiness dimensions.
    All readiness indicators were aggregated at the school level. 
    Overall readiness represents the composite index of the six institutional readiness dimensions.
    \end{flushleft}
\end{figure}

Students’ AI literacy exhibited substantial variability both within and between vocational institutions nationwide. As shown in Panels A and B of Figure~\ref{fig:school_readiness_descriptives}, the overall mean student AI literacy score was 3.008 (SD = 1.240), indicating considerable dispersion in AI-related competencies. When aggregated at the school level, mean student literacy scores ranged from 1.48 to 4.00 (median = 2.99), suggesting meaningful between-school differences in average performance. In addition, within-school variability differed across institutions, with school-specific standard deviations ranging from 0.24 to 1.67 (median = 1.26). This pattern indicates that schools varied not only in overall literacy levels but also in the degree of within-school performance dispersion.

School-level AI readiness demonstrated moderate dispersion across institutions and showed systematic associations with aggregated teacher mechanisms. As illustrated in Panels C-E of Figure~\ref{fig:school_readiness_descriptives}, readiness dimensions displayed variability consistent with institutional heterogeneity. Spearman correlations indicated that overall school AI readiness was positively associated with teachers’ perceived positive effects of AI ($\rho = 0.12$, $p < 0.001$), perceived negative effects ($\rho = 0.10$, $p < 0.001$), perceived difficulty of AI use ($\rho = 0.19$, $p < 0.001$), cognitive beliefs about AI ($\rho = 0.15$, $p < 0.001$), perceived workload burden ($\rho = 0.14$, $p < 0.001$), and perceived social support ($\rho = 0.19$, $p < 0.001$). In contrast, readiness was negatively associated with teachers’ satisfaction with AI use in teaching contexts ($\rho = -0.18$, $p < 0.001$). Although these correlations were modest in magnitude, their consistent direction suggests systematic co-variation between institutional readiness conditions and shared teacher-level perceptions, providing descriptive support for subsequent mediation analyses.

Preliminary multilevel modelling confirmed the presence of non-trivial between-school variance in student AI literacy. The unconditional model yielded an intraclass correlation coefficient (ICC) of 0.0118, indicating that 1.18\% of the total variance in student AI literacy was attributable to between-school differences. While modest, this magnitude is common in large-scale educational datasets and justifies multilevel modelling when theoretically meaningful higher-level predictors are examined \citep{hox2017multilevel}. Inclusion of school-level structural covariates reduced between-school variance by 15.7\%, indicating that regional and institutional characteristics accounted for a meaningful portion of observed school-level heterogeneity.

\subsection{School-level AI readiness and students’ AI literacy (H1)}

\begin{figure}
    \centering
    \includegraphics[width=0.75\linewidth]{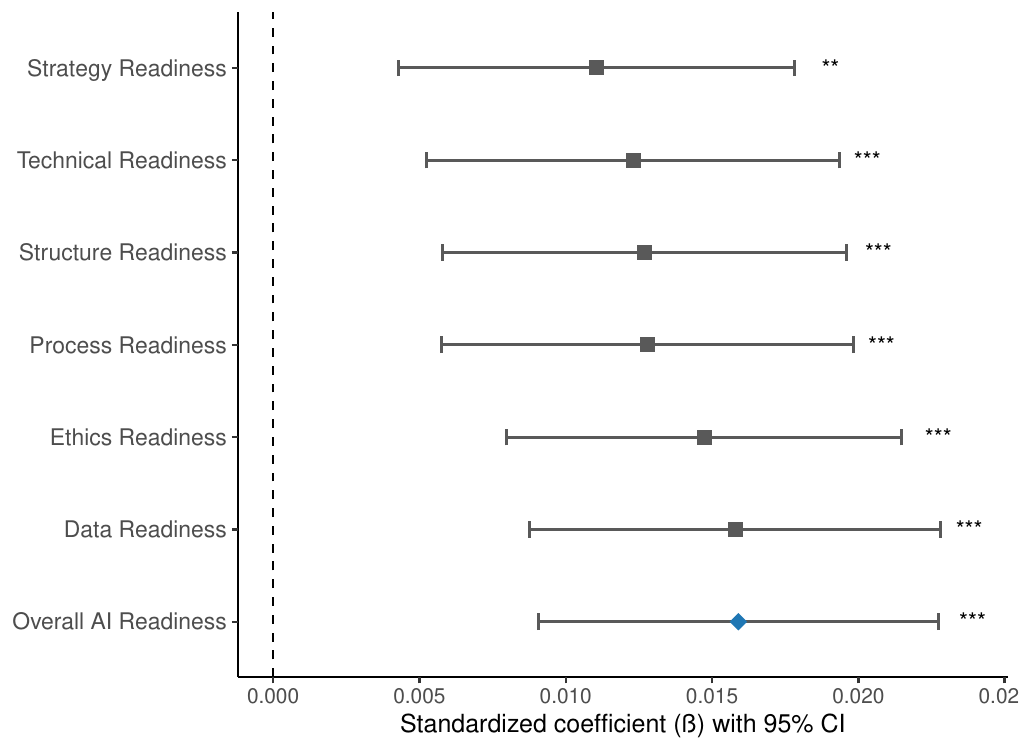}
    \caption{Associations between simultaneous school-level AI readiness dimensions and student AI literacy}
    \label{fig:Differentialimpactsofsimultaneous}
    \begin{flushleft}
    \footnotesize
    \textit{Note.} Squares represent standardised regression coefficients ($\beta$), and horizontal 
    lines indicate 95\% confidence intervals estimated from linear mixed-effects models with random 
    intercepts at the school level. Blue filled squares denote statistically significant effects 
    ($p < 0.05$), whereas grey open squares indicate non-significant effects. The dashed vertical 
    line marks the null association ($\beta = 0$).
    \end{flushleft}
\end{figure}

School-level AI readiness was positively associated with student AI literacy after adjusting for institutional structural characteristics. As shown in Figure~\ref{fig:Differentialimpactsofsimultaneous}, overall school AI readiness exhibited a statistically significant association with student AI literacy ($\beta = 0.016$, $t(df = 948.423) = 4.564$, $p < 0.001$) after controlling for economic-administrative region, policy status, and institutional type. This result indicates that vocational institutions with higher levels of organisational AI preparedness tended to demonstrate higher average levels of student AI literacy, independent of baseline structural differences.

When examined separately, all six readiness dimensions showed positive and statistically significant associations with student AI literacy. Specifically, strategy readiness ($\beta = 0.011$, $t(df = 949.756) = 3.199$, $p = 0.001$), organisational structure readiness ($\beta = 0.013$, $t(df = 941.653) = 3.598$, $p < 0.001$), process readiness ($\beta = 0.013$, $t(df = 948.233) = 3.566$, $p < 0.001$), technical readiness ($\beta = 0.012$, $t(df = 933.146) = 3.415$, $p < 0.001$), data readiness ($\beta = 0.016$, $t(df = 942.612) = 4.407$, $p < 0.001$), and ethics readiness ($\beta = 0.015$, $t(df = 953.607) = 4.282$, $p < 0.001$) each demonstrated consistent positive associations. Among these, data readiness showed the numerically largest standardised coefficient. Across models, regional indicators and institutional type (higher vs.\ secondary vocational) were consistently associated with student AI literacy, whereas policy status did not exhibit independent effects after covariate adjustment. Taken together, these findings support H1a and indicate that institutional AI readiness, both overall and across its dimensions, is positively related to student AI literacy when considered independently.

When the six readiness dimensions were entered simultaneously, individual coefficients were substantially attenuated and no longer reached conventional levels of statistical significance. Under this shared-variance specification, data readiness retained the numerically largest coefficient ($\beta = 0.0087$, $SE = 0.0053$, $t(df = 933.334) = 1.645$, $p = 0.10$), although the effect was not statistically significant. Adjusted generalised variance inflation factors (GVIF$^{1/(2df)}$) ranged from 2.16 to 3.96, remaining below commonly cited thresholds for severe multicollinearity (e.g., GVIF < 5). This pattern indicates moderate shared variance among readiness components, which is conceptually consistent with their interrelated organisational nature. Accordingly, the attenuation observed in the simultaneous model reflects overlapping institutional capacities rather than the absence of meaningful associations. Overall, these findings suggest that school AI readiness operates as an integrated configuration of mutually reinforcing components rather than as a set of independent, orthogonal predictors.

\subsection{Cross-level mediation of teacher-level mechanisms (H2)}

\begin{figure}
    \centering
    \includegraphics[width=\linewidth]{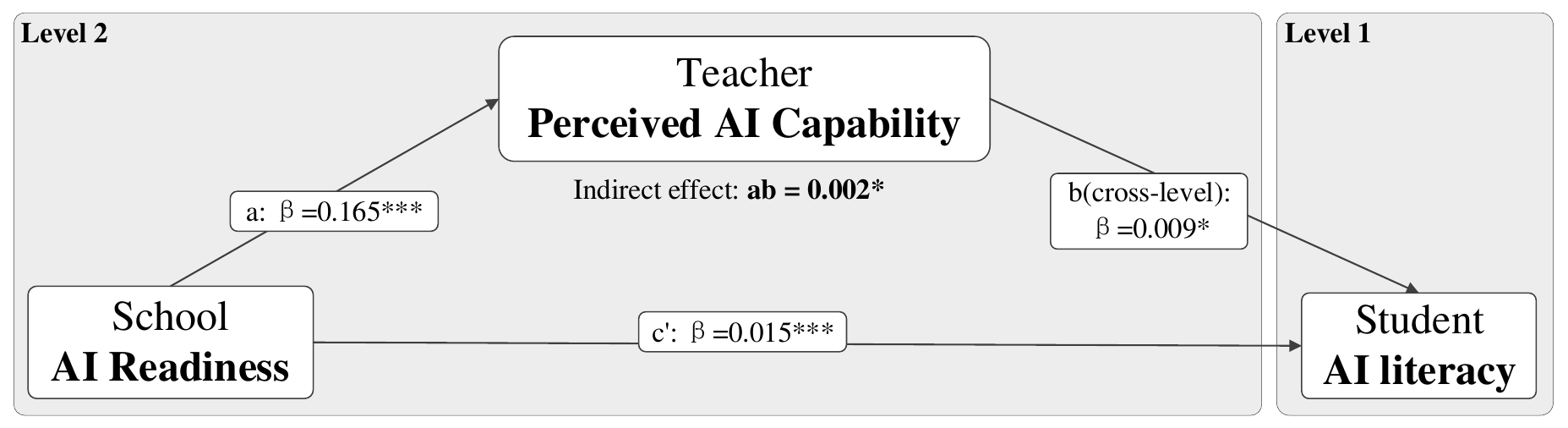}
    \caption{Cross-level mediation pathways linking overall school AI readiness, aggregated teacher mechanisms, and student AI literacy}
    \label{fig:cross_level_mediation}
    \begin{flushleft}
    \footnotesize
    \textit{Note.} Solid arrows indicate statistically significant paths. Dashed arrows represent 
    the direct association after accounting for mediators. Values are standardised regression 
    coefficients ($\beta$). Indirect effect ($a \times b$) is reported. 
    *$p$ < 0.05, **$p$ < 0.01, ***$p$ < 0.001.
    \end{flushleft}
\end{figure}

School-level AI readiness was systematically associated with multiple aggregated teacher-level mechanisms, establishing the first stage ($a$ paths) of the mediation model. As illustrated in Figure~\ref{fig:cross_level_mediation}, overall school AI readiness was positively associated with teachers’ perceived AI capability ($\beta = 0.165$, $SE = 0.033$, $t(df = 999) = 5.003$, $p < 0.001$), perceived positive instructional effects of AI ($\beta = 0.176$, $SE = 0.033$, $t(df = 999) = 5.308$, $p < 0.001$), perceived difficulty of using AI ($\beta = 0.208$, $SE = 0.033$, $t(df = 999) = 6.267$, $p < 0.001$), and perceived social support for AI-enabled teaching ($\beta = 0.185$, $SE = 0.033$, $t(df = 999) = 5.547$, $p < 0.001$). In contrast, readiness was negatively associated with teachers’ satisfaction with AI use in teaching ($\beta = -0.090$, $SE = 0.033$, $t(df = 999) = -2.754$, $p < 0.001$). These results indicate that institutional AI readiness is reflected in differentiated teacher-level perceptions, including capability beliefs, perceived instructional affordances, implementation demands, and support conditions.

Only selected teacher mechanisms demonstrated statistically significant cross-level associations with student AI literacy when entered simultaneously in the mediation model ($b$ paths). Among the examined constructs, teachers’ perceived AI capability was positively associated with student AI literacy after controlling for school AI readiness and structural covariates ($\beta = 0.009$, $SE = 0.004$, $t(df = 286{,}418) = 2.333$, $p < 0.001$). Other perception-based indicators exhibited weaker or statistically inconsistent associations in the full model. This pattern suggests that collective professional capability, rather than general evaluative attitudes or satisfaction, constitutes the most stable teacher-level predictor of student AI literacy in the cross-level framework.

Formal mediation testing confirmed a statistically significant indirect pathway linking school AI readiness to student AI literacy via teachers’ perceived AI capability. The estimated indirect effect was significant using the Delta method ($ab = 0.0015$, $z = 2.114$, $p = 0.034$) with a 95\% confidence interval excluding zero (CI = [0.00011,\ 0.00300]). Monte Carlo simulation further yielded a 95\% confidence interval of [0.00024,\ 0.00318], corroborating the robustness of the mediation effect. Importantly, the direct association between school AI readiness and student AI literacy remained statistically significant after accounting for the mediator ($c' = 0.015$, $SE = 0.004$, $t(df = 286{,}418) = 3.806$, $p < 0.001$), indicating partial mediation. Collectively, these findings support H2 and identify teachers’ perceived AI capability as a key organisational transmission mechanism through which institutional readiness is associated with student AI literacy outcomes.

\subsection{Boundary conditions and robustness of cross-level effects (H3)}

The mediation pathway linking school AI readiness, aggregated teacher capability, and student AI literacy remained robust under an aggregated school-level (2--2--2) specification. As shown in Table~\ref{tab:mediation_222}, school AI readiness was positively associated with teachers’ perceived AI capability ($\beta_a = 0.165$, $SE = 0.033$, $t(df = 999) = 5.003$, $p < 0.001$), consistent with the primary cross-level model. In turn, aggregated teacher capability was positively associated with aggregated student AI literacy ($\beta_b = 0.0136$, $SE = 0.0040$, $t(df = 996) = 3.372$, $p < 0.001$). The direct association between school AI readiness and student AI literacy remained statistically significant after accounting for the mediator ($\beta_{c'} = 0.0152$, $SE = 0.0040$, $t(df = 998) = 3.81$, $p < 0.001$), indicating partial mediation at the aggregated level. The indirect effect was statistically significant using the Delta method ($ab = 0.00225$, $z = 2.796$, $p = 0.005$, CI = [0.00067,\ 0.00383]) and remained significant under Monte Carlo simulation (mean $ab = 0.00226$, CI = [0.00083,\ 0.00404]). The replication of the mediation structure in a purely Level-2 framework provides convergent evidence that the capability-based transmission mechanism is not an artefact of cross-level modelling.

\begin{table}
\centering
\caption{Aggregated mediation results (2--2--2 model)}
\label{tab:mediation_222}
\begin{tabular}{p{4.5cm}ccccc}
\toprule
Path & $\beta$ & SE & Test statistic & df & $p$ / CI \\
\midrule
$a$: School AI readiness $\rightarrow$ teacher-perceived AI capability & 0.165 & 0.033 & $t = 5.00$ & 999 & $< .001$ \\
$b$: Teacher-perceived AI capability $\rightarrow$ Student AI literacy & 0.0136 & 0.0040 & $t = 3.37$ & 996 & $< .001$ \\
$c'$: Direct effect (School AI readiness $\rightarrow$ Student AI literacy) & 0.0152 & 0.0040 & $t = 3.81$ & 998 & $< .001$ \\
\midrule
Indirect effect ($ab$) -- Delta method & 0.00225 & 0.00081 & $z = 2.80$ &  & $=.005$ \\
Indirect effect ($ab$) -- Bootstrap & 0.00226 &  & -- & -- & CI [0.00083, 0.00404] \\
\bottomrule
\end{tabular}

\vspace{2mm}
\footnotesize
\raggedright
\textit{Note}. Bootstrap confidence intervals were estimated using 20,000 Monte Carlo resamples. All coefficients are standardised.
\end{table}

\begin{figure}
    \centering
    \includegraphics[width=0.75\linewidth]{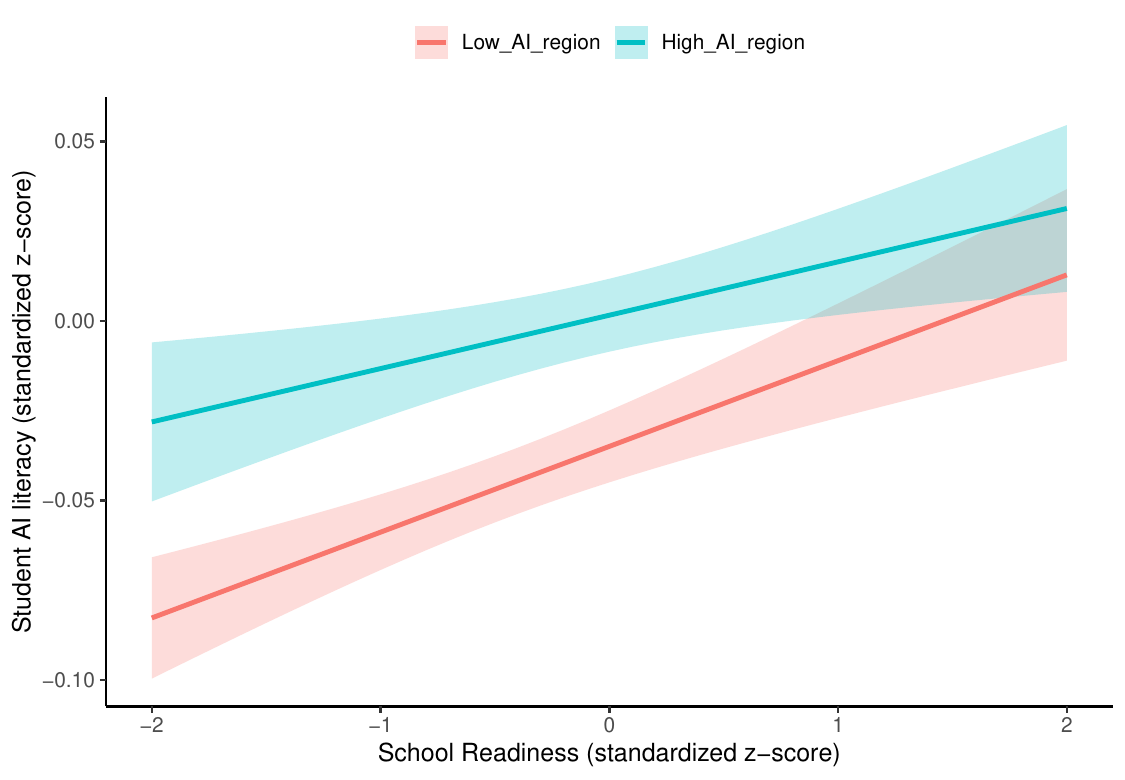}
    \caption{Associations between school AI readiness and student AI literacy across regional AI contexts}
    \label{fig:acrossregionalAIcontexts}
    \textit{Note.} Lines represent model-predicted values of student AI literacy across levels of school AI readiness in low-AI and high-AI regions. All variables are standardised (z-scores). Shaded areas indicate 95\% confidence intervals.
\end{figure}

Regional AI development exerted an independent contextual influence on student AI literacy but did not moderate the readiness--literacy association. In multilevel models including regional AI development (high vs.\ low classification), school AI readiness remained a significant positive predictor of student AI literacy ($\beta = 0.024$, $SE = 0.005$, $t(df = 1{,}006) = 5.171$, $p < 0.001$). Students in high AI-development regions demonstrated higher overall AI literacy levels ($\beta = 0.037$, $SE = 0.007$, $t(df = 9{,}349) = 5.022$, $p < 0.001$), indicating a contextual advantage. However, the interaction between school AI readiness and regional AI development was not statistically significant ($\beta = -0.009$, $SE = 0.007$, $t(df = 9{,}547) = -1.301$, $p = 0.194$), suggesting that the strength of the readiness--literacy association did not differ systematically between high- and low-AI regions. Simple slope analyses showed a statistically significant readiness effect in low-AI regions ($\beta = 0.016$, $SE = 0.005$, $t(df = 548.481) = 2.994$, $p < 0.01$) and a positive but non-significant slope in high-AI regions ($\beta = 0.006$, $SE = 0.004$, $t(df = 417.471) = 1.524$, $p = 0.13$); importantly, the difference between slopes was not statistically significant. Predicted values across ±2 SD of readiness are shown in Figure~\ref{fig:acrossregionalAIcontexts}. These findings indicate that while regional AI ecosystems elevate baseline literacy levels, they do not fundamentally alter the structural relationship between institutional readiness and student outcomes.

Province-level robustness checks further confirmed the stability of the readiness--literacy association across heterogeneous regional environments. As reported in Table~\ref{tab:moderation_industry}, school AI readiness remained a positive and statistically significant predictor of student AI literacy ($\beta = 0.014$, $SE = 0.004$, $t(df = 940.198) = 3.847$, $p < 0.001$) after controlling for contextual covariates. Regional AI industry development exhibited an independent positive association with student AI literacy ($\beta = 0.010$, $SE = 0.004$, $t(df = 944.035) = 2.738$, $p < 0.01$). However, the interaction between readiness and regional AI industry development was not statistically significant ($\beta = -0.001$, $SE = 0.003$, $t(df = 962.023) = -0.323$, $p = 0.746$). Together, these analyses provide consistent evidence that although contextual AI development conditions are associated with overall literacy levels, they do not operate as boundary conditions that meaningfully modify the structural effect of school AI readiness.

\begin{table}[t]
\centering
\caption{Effects of school AI readiness and regional AI industry development}
\label{tab:moderation_industry}
\begin{tabular}{lccccc}
\toprule
Predictor & $\beta$ & SE & $t$ & $p$ & $df$\\
\midrule
Intercept & -0.029 & 0.012 & -2.33 & 0.020 & 911.107 \\
School AI readiness ($z$) & 0.014 & 0.004 & 3.85 & $<$.001 & 940.198 \\
Regional AI industry development ($z$) & 0.010 & 0.004 & 2.74 & 0.006 & 944.035 \\
Readiness $\times$ Regional AI development & -0.001 & 0.003 & -0.32 & 0.746 & 962.023 \\
\bottomrule
\end{tabular}

\vspace{2mm}
\footnotesize\textit{Note}. Coefficients are standardized. Estimates are from linear mixed-effects models with random intercepts for schools. Models adjusted for regional economic zone, policy status, and school type (secondary vs.\ higher vocational); covariate coefficients are omitted for brevity.
\end{table}

\section{Discussion}

This study examined how school-level AI readiness is associated with student AI literacy and whether this relationship operates through aggregated teacher mechanisms within a large-scale vocational education system. By testing a 2-2-1 cross-level mediation framework across more than two million students and over one thousand institutions, the study moves beyond descriptive readiness audits and provides mechanism-based evidence on how institutional AI capacity is translated into student-level competencies.

\subsection{Main Findings}

The first major finding (RQ1) is that school-level AI readiness is positively associated with student AI literacy, both overall and across individual readiness dimensions. After adjusting for institutional type, regional economic zone, and policy designation, higher institutional readiness corresponded to higher average student AI literacy. When examined separately, all six readiness dimensions, strategic, organisational, process, technical, data, and ethical, showed positive associations, whereas simultaneous modelling revealed attenuation due to shared variance. This pattern supports an ecological and configurational interpretation of readiness rather than a single-dimension effect. Prior literature has conceptualised AI readiness largely as an infrastructural or governance checklist \citep{johnk2021ready,educause2023ai}, but empirical linkage to student learning outcomes has been limited. By demonstrating population-scale associations with student AI literacy, this study advances readiness from a descriptive institutional condition to an empirically supported explanatory construct. The results further align with broader digital competence research indicating that systemic school capacity conditions the learning opportunities available to students \citep{warschauer2010new,scherer2019technology}. The unique contribution lies in quantifying this association at scale within vocational education and showing that readiness functions as an integrated organisational configuration.

The second major finding (RQ2) is that aggregated teacher-perceived AI capability partially mediates the relationship between institutional readiness and student AI literacy. Although school AI readiness was associated with multiple teacher-level perceptions, including effort expectancy, social support, and satisfaction, only aggregated teacher-perceived AI capability demonstrated a statistically significant cross-level pathway to student outcomes. This finding refines technology adoption models such as UTAUT \citep{venkatesh2003user} by distinguishing general attitudinal acceptance from instructional capability. Prior AI-in-education scholarship has emphasised teacher beliefs and intentions \citep{miao2024ai,shen2024school}, yet has rarely modelled how aggregated professional perceptions translate into student competencies at scale. By empirically identifying aggregated teacher-perceived AI capability as the operative transmission mechanism, the study extends organisational support theory \citep{Rhoades2002} into the AI domain and shifts analytical focus from individual adoption toward coordinated instructional enactment. Importantly, the persistence of a significant direct effect indicates partial mediation, suggesting that additional organisational mechanisms, such as curriculum alignment or leadership processes, may also contribute.

The third major finding (RQ3) is that the readiness-capability-literacy pathway demonstrates structural robustness across heterogeneous regional AI ecosystems. While regional AI development independently elevated baseline student literacy levels, it did not significantly moderate the association between institutional readiness and student outcomes. Furthermore, the mediation structure replicated under an aggregated 2--2--2 specification, strengthening confidence that the identified pathway is not an artefact of cross-level modelling. Educational technology literature frequently cautions against contextual fragility \citep{zawacki2019systematic}, particularly when findings derived from technologically advanced regions fail to generalise to under-resourced contexts. Within a nationally diverse vocational system, however, the organisational logic linking readiness, collective capability, and student literacy remained stable. This suggests that while macro-environmental AI maturity shapes overall competency baselines, the internal institutional mechanism translating readiness into instructional capability operates with relative consistency. The unique contribution of this finding lies in demonstrating cross-regional structural stability at scale rather than relying on single-context case studies.

\subsection{Theoretical and Practical Implications}

The current findings carry important theoretical implications for research on AI integration in education by demonstrating that institutional readiness operates as a multilevel ecological process rather than as a standalone infrastructural condition. Much of the existing AI-in-education literature remains centred on individual-level constructs, such as student competence, teacher intention, or perceived usefulness \citep{long2020what,celik2022promises}. By explicitly modelling a 2-2-1 architecture, this study shows that macro-level institutional capacity influences micro-level student outcomes through meso-level collective teacher capability. This extends ecological systems theory \citep{Bronfenbrenner1994} into the AI domain and empirically supports the argument that technological transformation in education must be understood as an organisational phenomenon. The identification of aggregated teacher-perceived AI capability as a mediating construct further refines organisational support theory \citep{Rhoades2002} by demonstrating how shared professional perceptions translate institutional investments into student-level competencies. In doing so, the study contributes a scalable, mechanism-based explanatory framework that moves beyond adoption intention models and toward system-level theory building in educational AI research.

The findings also have significant implications for institutional leadership and policy design in vocational education systems undergoing AI transformation. First, institutional AI readiness should be conceptualised as an integrated organisational condition that combines infrastructure, governance, data capacity, and ethical oversight with sustained investment in collective teacher capability. The attenuation observed among readiness dimensions under simultaneous modelling indicates that isolated technical upgrades are unlikely to yield durable learning gains unless embedded within coordinated institutional alignment. Second, professional development policies should prioritise building teachers’ shared instructional capability rather than focusing solely on awareness campaigns or attitudinal acceptance. The mediation results suggest that teachers’ perceived capacity to design and enact AI-supported instruction is a more critical lever than general satisfaction or perceived social support. Third, national and regional evaluation frameworks for AI readiness may benefit from incorporating explicit indicators of collective instructional competence alongside traditional infrastructure metrics. In vocational settings where AI literacy must translate into occupation-specific task performance, policy emphasis on human capital development appears essential for converting macro-level AI investment into measurable student outcomes.

More broadly, the robustness of the readiness-capability-literacy pathway across heterogeneous regional AI ecosystems suggests that institutional organisational mechanisms may provide a scalable foundation for AI-enabled educational reform. While macro-level AI industry development elevates baseline competency levels, it does not fundamentally alter the internal school-level transmission mechanism. This implies that even in regions with comparatively lower AI industrial maturity, strengthening institutional readiness and teacher capability may constitute a viable strategy for narrowing AI literacy disparities. For policymakers seeking equitable AI transformation, the findings underscore that scalable improvement may be achieved not only through external ecosystem enhancement but also through strengthening internal organisational coherence within schools.

\subsection{Limitations and Future Research}

Several limitations warrant consideration when interpreting these findings and point toward important directions for future research. First, the cross-sectional design limits causal inference; although the theoretical model specifies institutional readiness as preceding aggregated teacher capability and student literacy, the observed relationships remain associational. Longitudinal or panel data would allow examination of temporal sequencing and dynamic evolution of readiness and capability over time, while intervention-based studies targeting teacher capability development could more directly test causal pathways. Second, teacher mechanisms were operationalised using aggregated perceptual indicators rather than direct behavioural measures of AI-supported instructional practice; although aggregation is theoretically justified for modelling school-level climate constructs, future studies could incorporate classroom observations, platform log data, or curriculum artefacts to triangulate capability with enacted pedagogy. Third, while regional AI industry development was examined as a contextual moderator, additional boundary conditions, such as leadership practices, occupational specialisation, industry-school partnerships, and regulatory environments, may shape how institutional readiness is operationalised across systems, and cross-national comparative research would further test generalisability. Finally, the relatively modest between-school variance in student AI literacy indicates that substantial heterogeneity remains at the individual level; future research should explore cross-level interactions to determine whether institutional readiness differentially benefits student subgroups based on prior digital competence, socioeconomic background, or program track. Addressing these limitations through longitudinal, behavioural, comparative, and equity-oriented designs will strengthen understanding of how institutional AI transformation unfolds across ecological levels.

\section{Conclusion}

As vocational education systems confront rapid industrial digitalisation, this national multilevel study clarifies how institutional conditions are linked to student AI literacy through the organisational ecology of AI integration. The findings indicate that school AI readiness is associated with higher student AI literacy, but that this relationship becomes educationally meaningful primarily when readiness is accompanied by stronger aggregated teacher instructional capability. Rather than functioning as independent inputs, institutional infrastructure, governance, and professional capacity operate as an integrated configuration shaping the instructional opportunities through which students develop AI literacy. From a policy and leadership perspective, the results suggest that investments in AI infrastructure are unlikely to yield sustained learning gains unless they are aligned with systematic, job-embedded professional development that strengthens teachers’ collective capability to design and enact AI-supported vocational learning. Strengthening this alignment may provide a scalable pathway for improving AI literacy in vocational education and for narrowing regional disparities as AI reshapes occupational practice.


\section*{Acknowledgement}
This work was supported by the National Social Science Foundation of China Major Project in Education (No. VCA230011). The authors would like to thank Yan Zhao (Tsinghua University) for her assistance in coordinating communication with provincial authorities and participating schools during the administration and collection of the large-scale survey.

\bibliographystyle{cas-model2-names}

\bibliography{0_reference}



\appendix
\section{Appendix}

\setcounter{table}{0}
\renewcommand{\thetable}{A\arabic{table}}

\setcounter{figure}{0}
\renewcommand{\thefigure}{A\arabic{figure}}

\begin{table*}
\centering
\caption{Provincial Minimum Sampling Targets and Achieved Sample Sizes by Institution Type}
\label{tab:prov_sampling}

\begin{tabular}{p{2.5cm} p{2.5cm} p{2.5cm} p{2.5cm} p{2cm}}
\toprule
Province & Secondary VET Min & Secondary VET Sampled & Higher VET Min & Higher VET Sampled \\
\midrule
Anhui & 11 & 24 & 16 & 21 \\
Beijing & 4 & 8 & 6 & 6 \\
Fujian & 7 & 12 & 11 & 9 \\
Gansu & 4 & 3 & 6 & 3 \\
Guangdong & 13 & 16 & 20 & 13 \\
Guangxi & 8 & 8 & 11 & 4 \\
Guizhou & 7 & 11 & 11 & 13 \\
Hainan & 3 & 10 & 4 & 7 \\
Hebei & 10 & 80 & 15 & 31 \\
Henan & 16 & 11 & 24 & 23 \\
Heilongjiang & 6 & 29 & 8 & 25 \\
Hubei & 9 & 38 & 14 & 35 \\
Hunan & 12 & 8 & 18 & 17 \\
Jilin & 4 & 10 & 7 & 6 \\
Jiangsu & 13 & 72 & 20 & 31 \\
Jiangxi & 10 & 0 & 14 & 10 \\
Liaoning & 7 & 20 & 11 & 17 \\
Inner Mongolia & 5 & 1 & 8 & 6 \\
Ningxia & 2 & 4 & 3 & 5 \\
Qinghai & 2 & 0 & 2 & 4 \\
Shandong & 13 & 66 & 19 & 52 \\
Shanxi & 7 & 14 & 11 & 10 \\
Shaanxi & 6 & 22 & 9 & 11 \\
Shanghai & 4 & 4 & 7 & 5 \\
Sichuan & 12 & 11 & 18 & 40 \\
Tianjin & 4 & 4 & 6 & 12 \\
Tibet & 1 & 0 & 1 & 1 \\
Xinjiang & 6 & 20 & 9 & 21 \\
Yunnan & 8 & 6 & 12 & 13 \\
Zhejiang & 7 & 12 & 11 & 19 \\
Chongqing & 7 & 9 & 10 & 8 \\
\bottomrule
\end{tabular}
\textit{Note}. Minimum sampling targets were predefined to ensure sufficient Level-2 clusters for multilevel modeling. Minor deviations occurred in a small number of provinces due to non-response or administrative constraints. 
\end{table*}

\begin{table*}
\centering
\caption{School Questionnaire Items Used in the Study (Bilingual)}
\label{tab:school_questionnaire_1}
\renewcommand{\arraystretch}{1.15}

\begin{tabular}{p{0.48\textwidth} p{0.48\textwidth}}
\toprule
\textbf{English Version} & \textbf{Chinese Original} \\
\midrule

\begin{minipage}[t]{0.48\textwidth}
\textbf{Q0.1. Has your school been selected for any of the following programs?}\\
(a) National Double-High Program (Top Vocational Colleges and Majors)\\
(b) Provincial Double-High Program\\
(c) National Double-Excellent Program\\
(d) Provincial Double-Excellent Program\\
(e) Other (please specify)
\end{minipage}
&
\begin{CJK}{UTF8}{gbsn}
\begin{minipage}[t]{0.48\textwidth}
\textbf{Q0.1. 贵校是否入选以下项目？}\\
(a) 国家级“双高计划”院校\\
(b) 省级“双高计划”院校等省级计划\\
(c) 国家级“双优计划”院校\\
(d) 省级“双优计划”院校等省级计划\\
(e) 其他（请注明）
\end{minipage}
\end{CJK}
\\
\midrule

\begin{minipage}[t]{0.48\textwidth}
\textbf{Q0.2. Which department do you belong to in the school?}\\
(a) Teacher Development or Teaching and Research Department\\
(b) Teaching Management Department\\
(c) Information Management Department\\
(d) Specialized department for promoting AI research or applications\\
(e) Other (please specify)
\end{minipage}
&
\begin{CJK}{UTF8}{gbsn}
\begin{minipage}[t]{0.48\textwidth}
\textbf{Q0.2. 您在学校中隶属于以下哪个部门？}\\
(a) 教师发展或教研部\\
(b) 教学管理部门\\
(c) 信息化管理部门\\
(d) 为推进AI研究或应用专门成立的部门\\
(e) 其他（请注明）
\end{minipage}
\end{CJK}
\\
\midrule

\multicolumn{2}{l}{\textbf{Dimension 1. Strategic Planning}}\\
\midrule
\begin{minipage}[t]{0.48\textwidth}
\textbf{Q1.1. How does your school respond to regional AI-related policies?}\\
(a) Unclear / No AI-related policies have been formulated in the region\\
(b) Understand the requirements of regional AI-related policies but have not taken action\\
(c) Considered regional policies when formulating planning schemes, but not comprehensively\\
(d) Formulated specific implementation plans to fully implement regional AI-related policies
\end{minipage}
&
\begin{CJK}{UTF8}{gbsn}
\begin{minipage}[t]{0.48\textwidth}
\textbf{您所在的学校对区域AI相关政策的响应情况如何: }\\
(a) 不清楚/本区域未制定AI相关政策\\
(b) 了解区域AI相关政策要求，未采取行动\\
(c) 在制定规划方案时考虑了区域相关政策，但不够全面\\
(d) 制定了具体的实施方案来全面落实区域AI相关政策
\end{minipage}
\end{CJK}
\\
\midrule

\begin{minipage}[t]{0.48\textwidth}
\textbf{Q1.2. How is your school doing in complying with laws and regulations to promote AI applications?}\\
(a) Do not know / Has not started yet\\
(b) Understand AI-related laws and regulations\\
(c) Formulated policies and guidelines for AI applications based on existing laws and regulations, but not comprehensively\\
(d) Strengthened publicity and training across the school, and supervised the entire process of AI applications in accordance with laws and regulations
\end{minipage}
&
\begin{CJK}{UTF8}{gbsn}
\begin{minipage}[t]{0.48\textwidth}
\textbf{Q1.2. 您所在的学校为推进人工智能应用，在依法依规方面的情况如何？}\\
(a) 不知道 / 尚未开始\\
(b) 了解AI相关法律法规\\
(c) 依据现有法律法规制定AI应用的政策和指南，但不够全面\\
(d) 全校范围内加强对现有法律法规的宣传和培训，并依法依规监管AI应用的全过程
\end{minipage}
\end{CJK}
\\
\midrule
\begin{minipage}[t]{0.48\textwidth}
\textbf{Q1.3. How has your school incorporated GenAI applications into the overall school plan?}\\
(a) Do not know / Has not started yet\\
(b) Recognize the importance of new technologies, but have not formed a clear development plan\\
(c) Formed a clear development plan related to GenAI applications, but limited to scenarios\\
(d) Formulated a comprehensive development plan related to GenAI applications that is consistent with the overall plan
\end{minipage}
&
\begin{CJK}{UTF8}{gbsn}
\begin{minipage}[t]{0.48\textwidth}
\textbf{Q1.3. 您所在的学校将GenAI应用纳入学校整体规划的情况如何？}\\
(a) 不知道 / 尚未开始\\
(b) 意识到新技术的重要性，但尚未形成明确的发展规划\\
(c) 形成明确的GenAI应用相关发展规划，但只局限于个别应用场景\\
(d) 制定了全面的GenAI应用相关发展规划，并与整体规划保持一致性
\end{minipage}
\end{CJK}
\\

\bottomrule

\end{tabular}
\end{table*}

\begin{table*}
\centering
\caption{School Questionnaire Items Used in the Study (Bilingual)\textit{(continued)}} 
\label{tab:school_questionnaire_2}
\renewcommand{\arraystretch}{1.15}

\begin{tabular}{p{0.48\textwidth} p{0.48\textwidth}}
\toprule
\textbf{English Version} & \textbf{Chinese Original} \\
\midrule

\begin{minipage}[t]{0.48\textwidth}
\textbf{Q1.4. How has your school formulated policies to promote AI applications?}\\
(a) Do not know / Has not started yet\\
(b) Recognize the importance of formulating relevant policies but have not formulated them\\
(c) Formulated relevant policies to promote AI applications, but not comprehensively, or included them in other informatization-related policies without dedicated policies\\
(d) Established full-scenario and whole-process relevant policies and promoted across the school
\end{minipage}
&
\begin{CJK}{UTF8}{gbsn}
\begin{minipage}[t]{0.48\textwidth}
\textbf{Q1.4. 您所在的学校为推进人工智能应用制定政策的情况如何？}\\
(a) 不知道 / 尚未开始\\
(b) 认识到制定相关政策的重要性，但尚未制定\\
(c) 已制定推进人工智能应用的相关政策，但还不够全面，或在其他信息化相关政策中有涉及但无专门的政策\\
(d) 已建立全场景、全过程的相关政策，并在全校进行了推广
\end{minipage}
\end{CJK}
\\
\midrule
\multicolumn{2}{l}{\textbf{Dimension 2. Organizational Structure}}\\
\midrule

\begin{minipage}[t]{0.48\textwidth}
\textbf{Q2.1. How adaptable is your school's organizational structure in promoting AI applications?}\\
(a) No intention to establish a leading organization, and no organizational restructuring for AI applications has been carried out\\
(b) Planning to establish a leading organization, and adaptive restructuring has just started\\
(c) A leading organization has been established, but the structure only meets basic needs and the overall effectiveness remains limited\\
(d) Established leading organization, and the structure is fully adapted (or no further adjustment is needed) to support AI application development
\end{minipage}
&
\begin{CJK}{UTF8}{gbsn}
\begin{minipage}[t]{0.48\textwidth}
\textbf{Q2.1. 您认为所在学校推动人工智能应用，在组织机构方面的适应性如何？}\\
(a) 目前既无设立牵头机构的意向，也未开展针对人工智能应用的组织结构调整\\
(b) 正在规划设立牵头机构，同时组织结构的适应性调整工作刚刚起步\\
(c) 已设立牵头机构，但组织结构仅能满足基本需求，整体推进效果仍有限\\
(d) 已成立专门牵头机构，且组织结构已充分适应（甚至无需额外调整）以支持人工智能应用的发展
\end{minipage}
\end{CJK}
\\
\midrule

\begin{minipage}[t]{0.48\textwidth}
\textbf{Q2.2. How is your school doing in cross-departmental collaboration to promote AI applications?}\\
(a) Do not know / Has not started yet\\
(b) Recognize the importance of cross-departmental and multi-stakeholder collaboration\\
(c) Established a cross-departmental team to coordinate work, but there is some overlap and conflict with the original departments\\
(d) Established an efficient cross-departmental team and optimized the division of functions to strengthen collaboration
\end{minipage}
&
\begin{CJK}{UTF8}{gbsn}
\begin{minipage}[t]{0.48\textwidth}
\textbf{Q2.2. 您所在的学校为推进人工智能应用，在跨部门协同方面的情况如何？}\\
(a) 不知道/尚未开始\\
(b) 意识到跨部门多元协同对于推进人工智能应用的重要性\\
(c) 建立了跨部门团队，统一协调团队工作，但与原部门工作存在一定的交叉和冲突\\
(d) 建立了完善高效的跨部门团队，并优化了职能划分，以加强部门协同
\end{minipage}
\end{CJK}
\\

\bottomrule

\end{tabular}
\end{table*}

\begin{table*}
\centering
\caption{School Questionnaire Items Used in the Study (Bilingual)\textit{(continued)}} 
\label{tab:school_questionnaire_3}
\renewcommand{\arraystretch}{1.15}

\begin{tabular}{p{0.48\textwidth} p{0.48\textwidth}}
\toprule
\textbf{English Version} & \textbf{Chinese Original} \\
\midrule

\begin{minipage}[t]{0.48\textwidth}
\textbf{Q2.3. How is your school doing in terms of leadership to promote AI applications?}\\
(a) School leaders have a negative attitude toward promoting new technologies such as GenAI\\
(b) Unclear / School leaders have not expressed opinions on new technologies\\
(c) School leaders recognize the importance of new technologies for school development but have not taken action\\
(d) School leaders have a strong desire to promote new technologies and have taken action\\
(e) School leaders have clear ideas and plans and have fostered institution-wide culture and consensus through various channels
\end{minipage}
&
\begin{CJK}{UTF8}{gbsn}
\begin{minipage}[t]{0.48\textwidth}
\textbf{Q2.3. 您所在的学校为推进人工智能应用，在领导力方面的情况如何？}\\
(a) 领导层对推动GenAI等新技术应用的态度不积极\\
(b) 不清楚/领导层尚未对新技术表达过看法\\
(c) 领导层认识到新技术对学校发展的重要性，但尚未有所行动\\
(d) 领导层有迫切推进新技术应用的想法，并已有所行动\\
(e) 领导层对推进新技术的应用有清晰的想法和规划，并通过多种途径将推进新技术发展为学校的整体文化和共识
\end{minipage}
\end{CJK}
\\
\midrule

\begin{minipage}[t]{0.48\textwidth}
\textbf{Q2.4. How accepting are your department leaders of GenAI?}\\
(a) Department leaders have a negative attitude toward promoting new technologies such as GenAI\\
(b) Unclear / Department leaders have not expressed opinions on new technologies\\
(c) Some department leaders recognize the importance of new technologies for departmental development\\
(d) Some department leaders have a strong desire to promote new technologies\\
(e) Some department leaders have specific ideas and plans for promoting new technologies
\end{minipage}
&
\begin{CJK}{UTF8}{gbsn}
\begin{minipage}[t]{0.48\textwidth}
\textbf{Q2.4. 您所在学校的部门领导对GenAI的接受程度如何？}\\
(a) 对推动GenAI等新技术应用的态度不积极\\
(b) 不清楚/部门领导尚未对新技术表达过看法\\
(c) 部分部门领导认识到新技术对部门发展的重要性\\
(d) 部分部门领导有迫切推进新技术应用的想法\\
(e) 部分部门领导对推进新技术的应用有了具体的想法和规划
\end{minipage}
\end{CJK}
\\
\midrule

\begin{minipage}[t]{0.48\textwidth}
\textbf{Q2.5. How is your school doing in assessing GenAI-related literacy?}\\
Options: \\
(a) Unclear/Not started; \\
(b) Preliminary assessment initiated (system incomplete); \\
(c) Relatively sound system (limited scope and depth); \\
(d) Mature and comprehensive assessment system)\\
Items:\\
- Teacher literacy assessment \\
- Management and service personnel literacy assessment\\
- Student literacy assessment
\end{minipage}
&
\begin{CJK}{UTF8}{gbsn}
\begin{minipage}[t]{0.48\textwidth}
\textbf{Q2.5. 您所在的学校在GenAI相关素养评估方面的情况如何？}\\
选项：\\
(a) 不知道/尚未开始；\\
(b) 已开始初步评估，但体系尚不完善；\\
(c) 有较为完善的评估体系，但范围和深度有待加深；\\
(d) 有成熟的评估体系，能全面、系统地评估\\
题目：\\
- 教师素养评估\\
- 管理和服务人员素养评估\\
- 学生素养评估
\end{minipage}
\end{CJK}
\\

\bottomrule

\end{tabular}
\end{table*}

\begin{table*}
\centering
\caption{School Questionnaire Items Used in the Study (Bilingual)\textit{(continued)}} 
\label{tab:school_questionnaire_4}
\renewcommand{\arraystretch}{1.15}

\begin{tabular}{p{0.48\textwidth} p{0.48\textwidth}}
\toprule
\textbf{English Version} & \textbf{Chinese Original} \\
\midrule

\begin{minipage}[t]{0.48\textwidth}
\textbf{Q2.6. How is your school doing in developing GenAI-related literacy? (Options: (1) Unclear/Not started; (2) Considered important but no specific plan; (3) Plans being formulated and gradually implemented; (4) Highly prioritized with clear plans and sustained promotion)}\\
Teacher literacy development\\
Management and service personnel literacy development\\
Student literacy development
\end{minipage}
&
\begin{CJK}{UTF8}{gbsn}
\begin{minipage}[t]{0.48\textwidth}
\textbf{Q2.6. 您所在的学校在GenAI相关素养发展方面的情况如何？(选项：1 不知道/尚未开始；2 重视，但尚未制定具体发展计划；3 正制定素养发展计划，并逐步实施；4 高度重视，已有明确的素养发展计划并持续推动)}\\
教师素养发展\\
管理和服务人员素养发展\\
学生素养发展
\end{minipage}
\end{CJK}
\\
\midrule

\begin{minipage}[t]{0.48\textwidth}
\textbf{Q2.7. How is your school doing in offering ``General AI Courses''?}\\
(a) Have not offered ``General AI Courses''\\
(b) Are planning to offer ``General AI Courses''\\
(c) Have offered, covering some majors\\
(d) Have offered, covering all majors
\end{minipage}
&
\begin{CJK}{UTF8}{gbsn}
\begin{minipage}[t]{0.48\textwidth}
\textbf{Q2.7. 您所在的学校“人工智能通识课程”开设情况如何？}\\
(a) 未开设“人工智能通识课程”\\
(b) 正在规划开设“人工智能通识课程”\\
(c) 已开设，覆盖部分专业\\
(d) 已开设，覆盖所有专业
\end{minipage}
\end{CJK}
\\
\midrule

\multicolumn{2}{l}{\textbf{Dimension 3. Process Guarantee}}\\
\midrule

\begin{minipage}[t]{0.48\textwidth}
\textbf{Q3.1. How is your school doing in terms of funding budget for promoting AI applications?}\\
(a) Unclear / No tendency to divide funding budgets for AI promotion yet\\
(b) Intend to divide funding budgets to promote AI applications\\
(c) Have formulated specific funding budget allocation plans\\
(d) Implemented the budget as planned and regularly reviewed the budget implementation status
\end{minipage}
&
\begin{CJK}{UTF8}{gbsn}
\begin{minipage}[t]{0.48\textwidth}
\textbf{Q3.1. 您所在的学校为推进人工智能应用，在有关经费预算方面的情况如何？}\\
(a) 不清楚/尚未有相关经费预算划分的倾向\\
(b) 有意向划分经费预算用以推进应用\\
(c) 已制定具体的相关经费预算分配方案\\
(d) 按计划执行预算，并定期审查预算执行情况
\end{minipage}
\end{CJK}
\\
\midrule

\begin{minipage}[t]{0.48\textwidth}
\textbf{Q3.2. How is your school doing in evaluating AI application projects?}\\
(a) Do not know / Has not started yet\\
(b) Began to recognize the importance of evaluation but have not implemented it\\
(c) Have conducted preliminary project evaluations but have not formed a sound evaluation system\\
(d) Have a relatively comprehensive and sound evaluation system, including input--output and sustainability evaluation
\end{minipage}
&
\begin{CJK}{UTF8}{gbsn}
\begin{minipage}[t]{0.48\textwidth}
\textbf{Q3.2. 您所在的学校在有关人工智能应用的项目评估方面的情况如何？}\\
(a) 不知道/尚未开始\\
(b) 开始意识到评估重要性，但尚未实施\\
(c) 已进行初步的项目评估，但尚未形成完善的评估体系\\
(d) 已有较为全面完善的评估体系，包括项目投入产出评估、可持续性评估等
\end{minipage}
\end{CJK}
\\
\midrule

\begin{minipage}[t]{0.48\textwidth}
\textbf{Q3.3. Please select the option ``Have taken preliminary actions but have not achieved obvious results'' for this question.(Attention check)}\\
(a) Unclear / Has not started yet\\
(b) Recognize the importance but have not taken action\\
(c) Have taken preliminary actions but have not achieved obvious results\\
(d) Have a relatively comprehensive action model and formed exemplary cases
\end{minipage}
&
\begin{CJK}{UTF8}{gbsn}
\begin{minipage}[t]{0.48\textwidth}
\textbf{Q3.3. 本题请选择“已经初步采取行动，但尚未有明显效果”选项。}\\
(a) 不知道/尚未开始\\
(b) 意识到重要性，但尚未采取行动\\
(c) 已经初步采取行动，但尚未有明显的效果\\
(d) 已经有较为全面完善的行动模式，并形成示范性案例
\end{minipage}
\end{CJK}
\\

\bottomrule

\end{tabular}
\end{table*}

\begin{table*}
\centering
\caption{School Questionnaire Items Used in the Study (Bilingual)\textit{(continued)}} 
\label{tab:school_questionnaire_5}
\renewcommand{\arraystretch}{1.15}

\begin{tabular}{p{0.48\textwidth} p{0.48\textwidth}}
\toprule
\textbf{English Version} & \textbf{Chinese Original} \\
\midrule

\begin{minipage}[t]{0.48\textwidth}
\textbf{Q3.4. How is your school doing in adopting large models and other intelligent technologies in education and teaching?}\\
(a) Unclear / Has not started yet (skip to Q3.5)\\
(b) Preliminary exploration in individual majors or courses without school-level coordination\\
(c) Conducted pilots in individual majors or courses\\
(d) Formed exemplary cases
\end{minipage}
&
\begin{CJK}{UTF8}{gbsn}
\begin{minipage}[t]{0.48\textwidth}
\textbf{Q3.4. 您所在的学校在教育教学方面采用大模型及其他智能技术的情况如何？}\\
(a) 不知道/尚未开始（请跳至Q3.5）\\
(b) 在个别专业或课程自发开展初步探索，但学校层面尚未有组织推进\\
(c) 已在个别专业或课程进行试点\\
(d) 已形成具有示范性的案例
\end{minipage}
\end{CJK}
\\
\midrule

\begin{minipage}[t]{0.48\textwidth}
\textbf{Q3.5. How is your school doing in adopting large models and other intelligent technologies in teacher and student development?}\\
(a) Unclear / Has not started yet (skip to Q3.6)\\
(b) Recognize the impact of GenAI but have not taken specific measures\\
(c) Preliminary exploration based on general-purpose large models\\
(d) Adopted dedicated large models or specialized platforms for deep integration
\end{minipage}
&
\begin{CJK}{UTF8}{gbsn}
\begin{minipage}[t]{0.48\textwidth}
\textbf{Q3.5. 您所在的学校在师生发展方面采用大模型及其他智能技术的情况如何？}\\
(a) 不知道/尚未开始（请跳至Q3.6）\\
(b) 意识到GenAI的重要影响，尚未采取具体措施\\
(c) 基于通用大模型开展初步探索\\
(d) 采用专用大模型或专用平台支持与业务深度融合
\end{minipage}
\end{CJK}
\\
\midrule

\begin{minipage}[t]{0.48\textwidth}
\textbf{Q3.6. How is your school doing in adopting large models and other intelligent technologies in industry--education integration?}\\
(a) Unclear / Has not started yet\\
(b) Recognize the impact of GenAI but have not taken specific measures\\
(c) Preliminary exploration based on general-purpose large models\\
(d) Adopted dedicated large models or specialized platforms for deep integration
\end{minipage}
&
\begin{CJK}{UTF8}{gbsn}
\begin{minipage}[t]{0.48\textwidth}
\textbf{Q3.6. 您所在的学校在产教融合方面采用大模型及其他智能技术的情况如何？}\\
(a) 不知道/尚未开始\\
(b) 意识到GenAI的重要影响，尚未采取具体措施\\
(c) 基于通用大模型开展初步探索\\
(d) 采用专用大模型或专用平台支持与业务深度融合
\end{minipage}
\end{CJK}
\\
\midrule

\begin{minipage}[t]{0.48\textwidth}
\textbf{Q3.7. How is your school doing in cooperating with suppliers and technical partners in AI applications?}\\
(a) Unclear / Has not started yet\\
(b) Initially contacted suppliers and partners but have not carried out cooperation\\
(c) Established cooperation relationships and formulated cooperation strategies\\
(d) Established stable cooperation relationships and promoted projects as planned
\end{minipage}
&
\begin{CJK}{UTF8}{gbsn}
\begin{minipage}[t]{0.48\textwidth}
\textbf{Q3.7. 您所在的学校在人工智能应用方面，与相关供应商和技术合作伙伴的合作情况如何？}\\
(a) 不知道/尚未开始\\
(b) 初步了解和接触供应商和技术合作伙伴，但尚未开展合作\\
(c) 与供应商、技术合作伙伴建立合作关系，并制定了合作策略\\
(d) 与供应商、技术合作伙伴建立稳定合作关系，按计划推进项目实施
\end{minipage}
\end{CJK}
\\

\bottomrule

\end{tabular}
\end{table*}

\begin{table*}
\centering
\caption{School Questionnaire Items Used in the Study (Bilingual)\textit{(continued)}} 
\label{tab:school_questionnaire_6}
\renewcommand{\arraystretch}{1.15}

\begin{tabular}{p{0.48\textwidth} p{0.48\textwidth}}
\toprule
\textbf{English Version} & \textbf{Chinese Original} \\
\midrule

\multicolumn{2}{l}{\textbf{Dimension 4. Ethics and Safety}}\\
\midrule

\begin{minipage}[t]{0.48\textwidth}
\textbf{Q4.1. How is your school doing in responding to AI-related risks?}\\
\textit{(Matrix single choice)}\\
Response scale:\\
(a) Unclear / Has not started\\
(b) Recognize the importance of risk control but have not taken action\\
(c) Publicize AI application risks to teachers and students and guide them to actively avoid risks\\
(d) Include violation penalties related to such risks in school regulations\\[4pt]
Items:\\
-- Technology dependence risks (e.g., excessive dependence leading to thinking inertia and reduced independent learning and innovation ability)\\
-- Misuse and abuse risks (e.g., cheating, plagiarism, copyright infringement, academic ethics violations)\\
-- Ethical and moral risks (e.g., algorithmic bias of GenAI leading to violations of laws, policies, and ethical norms)
\end{minipage}
&
\begin{CJK}{UTF8}{gbsn}
\begin{minipage}[t]{0.48\textwidth}
\textbf{Q4.1. 您所在的学校在有关AI风险应对方面的情况如何？}\\
（矩阵单选）\\
选项：\\
(a) 不知道/尚未开始\\
(b) 意识到相关风险控制的重要性，但尚未行动\\
(c) 向师生宣讲AI应用风险，引导师生主动规避风险\\
(d) 在相关规章制度中包含关于该风险的违规处罚规定\\[4pt]
项目：\\
-- 技术依赖风险（过度依赖可能导致思维惰性，降低自主学习能力和创新能力）\\
-- 乱用滥用风险（如作弊、抄袭、版权侵犯或学术伦理问题）\\
-- 伦理道德风险（如GenAI算法偏见引发的法律、政策和伦理风险）
\end{minipage}
\end{CJK}
\\
\midrule

\begin{minipage}[t]{0.48\textwidth}
\textbf{Q4.2. How is your school doing in terms of AI safety management (e.g., data security, algorithm security, identity and permission management)?}\\
(a) Unclear / Has not started yet\\
(b) Understand the necessity of AI safety management but have not taken specific measures\\
(c) Have formulated basic AI safety management measures, but coverage and depth are insufficient\\
(d) Established a sound AI safety management tracking and supervision system
\end{minipage}
&
\begin{CJK}{UTF8}{gbsn}
\begin{minipage}[t]{0.48\textwidth}
\textbf{Q4.2. 您所在的学校在有关AI安全管理（如数据安全、算法安全、身份与权限管理等）方面的情况如何？}\\
(a) 不知道/尚未开始\\
(b) 了解AI安全管理的必要性，但未采取具体措施\\
(c) 已制定基本AI安全管理措施，但覆盖面和深度不够全面\\
(d) 建立了完善的AI安全管理追踪和监管体系
\end{minipage}
\end{CJK}
\\
\midrule

\multicolumn{2}{l}{\textbf{Dimension 5. Data Preparation}}\\
\midrule

\begin{minipage}[t]{0.48\textwidth}
\textbf{Q5.1. How is your school doing in terms of data standardization?}\\
(a) Unclear / Has not started yet\\
(b) Recognize the importance of data standardization and data requirements but have not taken specific actions\\
(c) Formulated data standards for different types of data\\
(d) Formulated data evaluation indicators based on data specifications and data standards
\end{minipage}
&
\begin{CJK}{UTF8}{gbsn}
\begin{minipage}[t]{0.48\textwidth}
\textbf{Q5.1. 您所在学校在数据规范方面情况如何？}\\
(a) 不知道/尚未开始\\
(b) 意识到数据规范和数据要求的重要性，但尚无具体行动\\
(c) 针对不同类型的数据制定了数据标准\\
(d) 依据数据规范和数据标准制定了数据评估指标
\end{minipage}
\end{CJK}
\\

\bottomrule

\end{tabular}
\end{table*}

\begin{table*}
\centering
\caption{School Questionnaire Items Used in the Study (Bilingual)\textit{(continued)}} 
\label{tab:school_questionnaire_7}
\renewcommand{\arraystretch}{1.15}

\begin{tabular}{p{0.48\textwidth} p{0.48\textwidth}}
\toprule
\textbf{English Version} & \textbf{Chinese Original} \\
\midrule

\begin{minipage}[t]{0.48\textwidth}
\textbf{Q5.2. How is your school doing in terms of data integration and management?}\\
(a) Unclear / Has not integrated yet; data is scattered in different silos\\
(b) Recognize the importance of data integration and initially plan to establish a data center\\
(c) Built a data center and integrated most of the data; some silos may still exist\\
(d) Data in the data center is uniformly managed and access methods are provided
\end{minipage}
&
\begin{CJK}{UTF8}{gbsn}
\begin{minipage}[t]{0.48\textwidth}
\textbf{Q5.2. 您所在学校在数据集成和管理方面情况如何？}\\
(a) 不知道/尚未集成，数据分散在不同孤岛中\\
(b) 意识到数据集成的重要性，初步拟定建立数据中心\\
(c) 已建设数据中心并集成大部分数据，但仍可能存在部分孤岛\\
(d) 数据中心中的数据得到统一管理，并提供访问方式
\end{minipage}
\end{CJK}
\\
\midrule

\begin{minipage}[t]{0.48\textwidth}
\textbf{Q5.3. How is your school doing in establishing data usage mechanisms?}\\
(a) Unclear / Have not established clear data usage mechanisms yet\\
(b) Recognize the importance of building data usage mechanisms but have not taken specific actions\\
(c) Established data usage agreement mechanisms with basic access restrictions, but not yet perfect\\
(d) Established sound data usage agreement mechanisms that meet basic access needs of AI applications
\end{minipage}
&
\begin{CJK}{UTF8}{gbsn}
\begin{minipage}[t]{0.48\textwidth}
\textbf{Q5.3. 您所在学校在建立数据使用机制方面情况如何？}\\
(a) 不知道/尚未建立明确的数据使用机制\\
(b) 意识到构建数据使用机制的重要性，但尚无具体行动\\
(c) 建立数据使用协议机制，对数据访问形成一定约束，但仍不完善\\
(d) 建立完善的数据使用协议机制，满足人工智能应用的基本访问需求
\end{minipage}
\end{CJK}
\\
\midrule

\begin{minipage}[t]{0.48\textwidth}
\textbf{Q5.4. How is your school doing in preparing high-quality data?}\\
\textit{(Matrix single choice)}\\
Response scale:\\
(a) Unclear / Has not started\\
(b) Organized collection of relevant data\\
(c) Conduct data management based on data standardization and quality standards\\
(d) Conduct preprocessing such as annotation and structuring based on scenarios\\[4pt]
Items:\\
-- Teaching content data (e.g., digital textbooks, curriculum standards, teaching plans, evaluation standards)\\
-- Teaching process data (e.g., learning logs, evaluations, classroom feedback, reports, assignments, works)\\
-- Teaching resource data (e.g., video, text, simulation training resources)\\
-- Vocational field data (e.g., industry standards, enterprise cases, recruitment and employment data)\\
-- Other internal data (e.g., student files, teacher files, policy texts, management processes)\\
-- External data (e.g., national/regional policies, laws and regulations, research reports)
\end{minipage}
&
\begin{CJK}{UTF8}{gbsn}
\begin{minipage}[t]{0.48\textwidth}
\textbf{Q5.4. 您所在的学校在高质量数据准备方面的情况如何？}\\
（矩阵单选）\\
选项：\\
(a) 不知道/尚未开始\\
(b) 有组织地收集相关数据\\
(c) 基于数据规范性和质量标准开展数据管理\\
(d) 基于应用场景开展标注、结构化等预处理\\[4pt]
项目：\\
-- 教学内容数据（如数字教材、课程标准、教学方案、评测标准等）\\
-- 教学过程数据（如学习日志、评教、课堂反馈、实训/实验报告、作业和作品等）\\
-- 教学资源数据（如视频、文本、仿真实训资源等）\\
-- 职业领域数据（如行业规范、企业案例、招聘与就业数据等）\\
-- 内部档案数据（如学生档案、教师档案、政策文本、业务流程等）\\
-- 外部相关数据（如国家或区域政策文件、法律法规、研究报告等）
\end{minipage}
\end{CJK}
\\

\bottomrule

\end{tabular}
\end{table*}

\begin{table*}
\centering
\caption{School Questionnaire Items Used in the Study (Bilingual)\textit{(continued)}} 
\label{tab:school_questionnaire_8}
\renewcommand{\arraystretch}{1.15}

\begin{tabular}{p{0.48\textwidth} p{0.48\textwidth}}
\toprule
\textbf{English Version} & \textbf{Chinese Original} \\
\midrule

\multicolumn{2}{l}{\textbf{Dimension 6. Technical Environment}}\\
\midrule

\begin{minipage}[t]{0.48\textwidth}
\textbf{Q6.1. How is your school doing in terms of computing power resources?}\\
(a) Unclear / No computing power resources available\\
(b) Recognize the importance of computing power resources for promoting GenAI applications and intend to or have conducted preliminary exploration\\
(c) Use third-party computing power services or existing computing resources to support AI projects for teaching or research \textit{(Please specify GPU cluster FLOPS)}\\
(d) Expanded computing resources and established GPU infrastructure to meet current and future AI project needs \textit{(Please specify GPU cluster FLOPS)}
\end{minipage}
&
\begin{CJK}{UTF8}{gbsn}
\begin{minipage}[t]{0.48\textwidth}
\textbf{Q6.1. 您所在学校在算力资源方面情况如何？}\\
(a) 不知道/尚无任何算力资源\\
(b) 意识到算力资源对推进GenAI应用的重要性，已有意愿或初步探索\\
(c) 使用第三方算力服务或现有计算资源支持教学或科研AI项目（请填写GPU集群总算力量FLOPS）\\
(d) 拓展现有计算资源并建设GPU基础设施，满足当前及未来AI项目需求（请填写GPU集群总算力量FLOPS）
\end{minipage}
\end{CJK}
\\
\midrule

\begin{minipage}[t]{0.48\textwidth}
\textbf{Q6.2. Does your school have technical personnel competent for large model deployment and operation and maintenance?}\\
(a) Unclear / No\\
(b) Have personnel with basic understanding of relevant technologies\\
(c) Have personnel who have participated in relevant projects\\
(d) Can form a technical team to undertake related work
\end{minipage}
&
\begin{CJK}{UTF8}{gbsn}
\begin{minipage}[t]{0.48\textwidth}
\textbf{Q6.2. 您所在学校是否具备胜任大模型部署与运维的技术人员？}\\
(a) 不清楚/不具备\\
(b) 有了解相关技术的人员\\
(c) 有参与过相关项目的技术人员\\
(d) 可组建技术团队承担相关工作
\end{minipage}
\end{CJK}
\\
\midrule

\begin{minipage}[t]{0.48\textwidth}
\textbf{Q6.3. How is your school doing in terms of large model deployment and development?}\\
(a) Unclear / Has not started yet\\
(b) Recognize the supporting role of large models for teaching and learning but have not taken actions\\
(c) Locally deployed one or more base large models and can provide common intelligent tools\\
(d) Built dedicated large models providing professional tools for specific application scenarios
\end{minipage}
&
\begin{CJK}{UTF8}{gbsn}
\begin{minipage}[t]{0.48\textwidth}
\textbf{Q6.3. 您所在学校在大模型部署和开发方面的情况如何？}\\
(a) 不知道/尚未开始\\
(b) 意识到大模型对教与学的支持作用，但尚无具体行动\\
(c) 本地部署一个或多个基座大模型，提供常见智能工具\\
(d) 构建专用大模型，提供面向特定场景的专业智能工具
\end{minipage}
\end{CJK}
\\
\midrule

\begin{minipage}[t]{0.48\textwidth}
\textbf{Q6.4. How is the interoperability between existing digital campus systems and large models in your school?}\\
(a) Unclear / Has not started yet\\
(b) Recognize the importance of interoperability but have not taken actions\\
(c) Can customize integration interfaces according to application needs, but not yet perfect\\
(d) Formulated integration interface schemes and opened interfaces for large model integration
\end{minipage}
&
\begin{CJK}{UTF8}{gbsn}
\begin{minipage}[t]{0.48\textwidth}
\textbf{Q6.4. 您所在学校数字校园已有应用系统与大模型的互操作性情况如何？}\\
(a) 不知道/尚未开始\\
(b) 意识到互操作性的重要性，但尚无具体行动\\
(c) 可按应用场景需求定制集成接口，但尚不完善\\
(d) 制定并开放与大模型集成的接口方案
\end{minipage}
\end{CJK}
\\

\bottomrule

\end{tabular}
\end{table*}

\begin{table*}
\centering
\caption{School Questionnaire Items Used in the Study (Bilingual)\textit{(continued)}} 
\label{tab:school_questionnaire_9}
\renewcommand{\arraystretch}{1.15}

\begin{tabular}{p{0.48\textwidth} p{0.48\textwidth}}
\toprule
\textbf{English Version} & \textbf{Chinese Original} \\
\midrule

\begin{minipage}[t]{0.48\textwidth}
\textbf{Q6.5. How is your school doing in terms of intelligent agent development?}\\
(a) Unclear / Has not started yet\\
(b) Have purchased or used general intelligent agents developed by third-party platforms\\
(c) Entrusted third parties to customize intelligent agents based on school application needs and data\\
(d) Independently configure agents using development platforms to meet application needs
\end{minipage}
&
\begin{CJK}{UTF8}{gbsn}
\begin{minipage}[t]{0.48\textwidth}
\textbf{Q6.5. 您所在的学校在智能体开发方面情况如何？}\\
(a) 不知道/尚未开始\\
(b) 已购买或使用第三方平台开发的通用智能体\\
(c) 委托第三方基于学校场景需求和数据定制开发智能体\\
(d) 基于智能体开发平台自主灵活配置智能体，满足应用需求
\end{minipage}
\end{CJK}
\\

\bottomrule

\end{tabular}
\end{table*}

\begin{table*}
\centering
\caption{Teacher Questionnaire Items Used in the Study (Bilingual)} 
\label{tab:teacher_questionnaire_1}
\renewcommand{\arraystretch}{1.15}

\begin{tabular}{p{0.48\textwidth} p{0.48\textwidth}}
\toprule
\textbf{English Version} & \textbf{Chinese Original} \\
\midrule
\begin{minipage}[t]{0.48\textwidth}
\textbf{T0.1. What is your gender?}\\
(a) Male\\
(b) Female
\end{minipage}
&
\begin{CJK}{UTF8}{gbsn}
\begin{minipage}[t]{0.48\textwidth}
\textbf{T0.1. 您的性别为？}\\
(a) 男\\
(b) 女
\end{minipage}
\end{CJK}
\\
\midrule

\begin{minipage}[t]{0.48\textwidth}
\textbf{T0.2. How many years of teaching experience do you have?}\\
(a) Less than 1 year\\
(b) 1--3 years\\
(c) 4--10 years\\
(d) More than 10 years
\end{minipage}
&
\begin{CJK}{UTF8}{gbsn}
\begin{minipage}[t]{0.48\textwidth}
\textbf{T0.2. 您的教龄为？}\\
(a) 1年以下\\
(b) 1--3年\\
(c) 4--10年\\
(d) 10年以上
\end{minipage}
\end{CJK}
\\
\midrule

\begin{minipage}[t]{0.48\textwidth}
\textbf{T0.3. Which discipline category does your teaching major belong to?}\\
(a) General education courses (e.g., ideology and politics, Chinese, mathematics, English)\\
(b) Agriculture, Forestry, Animal Husbandry and Fishery\\
(c) Resources, Environment and Safety\\
(d) Energy, Power and Materials\\
(e) Biology and Chemical Engineering\\
(f) Civil Engineering and Architecture\\
(g) Water Conservancy\\
(h) Equipment Manufacturing\\
(i) Light Industry and Textile\\
(j) Transportation\\
(k) Electronic Information\\
(l) Food, Medicine and Grain\\
(m) Medical and Health\\
(n) Finance and Commerce\\
(o) Culture and Arts\\
(p) Education and Sports\\
(q) Journalism and Communication\\
(r) Tourism\\
(s) Public Security and Justice\\
(t) Public Management and Services\\
(u) Other (please specify)\\
(v) Currently not teaching
\end{minipage}
&
\begin{CJK}{UTF8}{gbsn}
\begin{minipage}[t]{0.48\textwidth}
\textbf{T0.3. 您任教专业所属的专业大类为？}\\
(a) 公共课（思政、语文、数学、英语等）\\
(b) 农林牧渔大类\\
(c) 资源环境与安全大类\\
(d) 能源动力与材料大类\\
(e) 生物与化工大类\\
(f) 土木建筑大类\\
(g) 水利大类\\
(h) 装备制造大类\\
(i) 轻工纺织大类\\
(j) 交通运输大类\\
(k) 电子信息大类\\
(l) 食品药品与粮食大类\\
(m) 医药卫生大类\\
(n) 财经商贸大类\\
(o) 文化艺术大类\\
(p) 教育与体育大类\\
(q) 新闻传播大类\\
(r) 旅游大类\\
(s) 公安和司法大类\\
(t) 公共管理与服务大类\\
(u) 其他（请注明）\\
(v) 目前不教课
\end{minipage}
\end{CJK}
\\

\bottomrule

\end{tabular}
\end{table*}

\begin{table*}
\centering
\caption{Teacher Questionnaire Items Used in the Study (Bilingual)\textit{(continued)}} 
\label{tab:teacher_questionnaire_2}
\renewcommand{\arraystretch}{1.15}

\begin{tabular}{p{0.48\textwidth} p{0.48\textwidth}}
\toprule
\textbf{English Version} & \textbf{Chinese Original} \\
\midrule

\multicolumn{2}{l}{\textbf{Dimension 1. Perceived Positive Instructional Effects}}\\
\midrule

\begin{minipage}[t]{0.48\textwidth}
\textbf{T1.1. I believe that using generative AI will have positive impacts on the following aspects.}\\
\textit{(Matrix single choice)}\\
Response scale:\\
(a) Strongly disagree\\
(b) Disagree\\
(c) Neutral\\
(d) Agree\\
(e) Strongly agree\\[4pt]
Items:\\
-- Facilitate personalized instruction\\
-- Improve my teaching competence\\
-- Reduce my teaching workload\\
-- Improve teaching effectiveness\\
-- Increase students' learning interest\\
-- Enhance students' engagement\\
-- Improve students' self-regulated learning ability\\
-- Enhance higher-order thinking skills (e.g., problem solving, critical thinking, creativity)
\end{minipage}
&
\begin{CJK}{UTF8}{gbsn}
\begin{minipage}[t]{0.48\textwidth}
\textbf{T1.1. 我认为使用生成式人工智能会对以下方面产生积极影响。}\\
（矩阵单选）\\
选项：\\
(a) 非常不同意\\
(b) 不太同意\\
(c) 一般\\
(d) 比较同意\\
(e) 非常同意\\[4pt]
项目：\\
-- 有助于开展个性化教学\\
-- 能够提升我的教学能力\\
-- 能够减轻我的教学工作量\\
-- 能够提高教学效果\\
-- 能够提升学生学习兴趣\\
-- 能够提高学生参与度\\
-- 能够提高学生自主学习能力\\
-- 能够加强高阶思维能力（如问题解决、批判性思维、创新性等）
\end{minipage}
\end{CJK}
\\
\midrule
\multicolumn{2}{l}{\textbf{Dimension 2. Perceived Negative Instructional Effects}}\\
\midrule
\begin{minipage}[t]{0.48\textwidth}
\textbf{T1.2. I believe that using generative AI will have negative impacts on the following aspects.}\\
\textit{(Matrix single choice)}\\
Response scale:\\
(a) Strongly disagree\\
(b) Disagree\\
(c) Neutral\\
(d) Agree\\
(e) Strongly agree\\[4pt]
Items:\\
-- Increase the risk of academic integrity issues\\
-- Incorrect information may mislead students\\
-- Weaken students' mastery of foundational knowledge\\
-- Reduce students' critical thinking ability\\
-- Hinder the improvement of students' learning ability\\
-- Increase teachers' dependence on AI and reduce teaching innovation\\
-- Reduce emotional interaction between teachers and students
\end{minipage}
&
\begin{CJK}{UTF8}{gbsn}
\begin{minipage}[t]{0.48\textwidth}
\textbf{T1.2. 我认为使用生成式人工智能会对以下方面产生消极影响。}\\
（矩阵单选）\\
选项：\\
(a) 非常不同意\\
(b) 不太同意\\
(c) 一般\\
(d) 比较同意\\
(e) 非常同意\\[4pt]
项目：\\
-- 难以避免学业诚信问题\\
-- 错误的知识信息可能对学生产生误导\\
-- 学生基础知识掌握不牢\\
-- 学生逐渐丧失批判性思维能力\\
-- 不利于学生提升学习能力\\
-- 教师容易产生依赖，进而削弱教学创新能力\\
-- 减少教师与学生之间的情感交流
\end{minipage}
\end{CJK}
\\

\bottomrule

\end{tabular}
\end{table*}

\begin{table*}
\centering
\caption{Teacher Questionnaire Items Used in the Study (Bilingual)\textit{(continued)}} 
\label{tab:teacher_questionnaire_3}
\renewcommand{\arraystretch}{1.15}

\begin{tabular}{p{0.48\textwidth} p{0.48\textwidth}}
\toprule
\textbf{English Version} & \textbf{Chinese Original} \\
\midrule

\multicolumn{2}{l}{\textbf{Dimension 3. Perceived Difficulty of Using AI}}\\
\midrule

\begin{minipage}[t]{0.48\textwidth}
\textbf{T1.3. Perceived difficulty of using generative AI}\\
Response scale:\\
(a) Strongly disagree\\
(b) Disagree\\
(c) Neutral\\
(d) Agree\\
(e) Strongly agree\\[4pt]
Items:\\
-- I need to spend a lot of time and effort learning how to use generative AI.\\
-- I think using generative AI is difficult.\\
-- I think integrating generative AI into teaching will increase additional workload.
\end{minipage}
&
\begin{CJK}{UTF8}{gbsn}
\begin{minipage}[t]{0.48\textwidth}
\textbf{T1.3. 感知使用难度（生成式人工智能）}\\
选项：\\
(a) 非常不同意\\
(b) 不太同意\\
(c) 一般\\
(d) 比较同意\\
(e) 非常同意\\[4pt]
项目：\\
-- 我需要花费大量时间和精力学习生成式人工智能的使用\\
-- 我认为使用生成式人工智能很困难\\
-- 我认为将生成式人工智能融入教学会增加额外工作量
\end{minipage}
\end{CJK}
\\
\midrule

\multicolumn{2}{l}{\textbf{Dimension 4. Satisfaction with AI Use in Teaching}}\\
\midrule

\begin{minipage}[t]{0.48\textwidth}
\textbf{T1.4. Satisfaction with AI use in teaching}\\
Response scale:\\
(a) Strongly disagree\\
(b) Disagree\\
(c) Neutral\\
(d) Agree\\
(e) Strongly agree\\[4pt]
Items:\\
-- I am satisfied with the instructional outcomes brought by using generative AI.\\
-- I find the process of using generative AI in teaching enjoyable and interesting.\\
-- Generative AI can effectively improve my work efficiency beyond teaching tasks.\\
-- I will continue to use generative AI in my teaching in the future.
\end{minipage}
&
\begin{CJK}{UTF8}{gbsn}
\begin{minipage}[t]{0.48\textwidth}
\textbf{T1.4. 教学中使用生成式人工智能的满意度}\\
选项：\\
(a) 非常不同意\\
(b) 不太同意\\
(c) 一般\\
(d) 比较同意\\
(e) 非常同意\\[4pt]
项目：\\
-- 我对使用生成式人工智能带来的教学结果是满意的\\
-- 我使用生成式人工智能开展教学的过程是愉悦、有趣的\\
-- 生成式人工智能能够有效提升我除教学外的工作效率\\
-- 我今后会持续使用生成式人工智能开展教学
\end{minipage}
\end{CJK}
\\
\midrule

\multicolumn{2}{l}{\textbf{Dimension 5. Perceived Social Support for AI-Enabled Teaching}}\\
\midrule

\begin{minipage}[t]{0.48\textwidth}
\textbf{T1.5. Perceived social support for AI-enabled teaching}\\
Response scale:\\
(a) Strongly disagree\\
(b) Disagree\\
(c) Neutral\\
(d) Agree\\
(e) Strongly agree\\[4pt]
Items:\\
-- National and school-level policies encourage me to use generative AI in teaching.\\
-- Leaders, colleagues, peers, and students around me encourage or expect me to use generative AI in teaching.
\end{minipage}
&
\begin{CJK}{UTF8}{gbsn}
\begin{minipage}[t]{0.48\textwidth}
\textbf{T1.5. 面向AI赋能教学的社会支持感知}\\
选项：\\
(a) 非常不同意\\
(b) 不太同意\\
(c) 一般\\
(d) 比较同意\\
(e) 非常同意\\[4pt]
项目：\\
-- 国家及学校政策鼓励我使用生成式人工智能开展教学\\
-- 我身边的领导、同事、同行及学生鼓励或期望我使用生成式人工智能开展教学
\end{minipage}
\end{CJK}
\\

\bottomrule

\end{tabular}
\end{table*}

\begin{table*}
\centering
\caption{Teacher Questionnaire Items Used in the Study (Bilingual)\textit{(continued)}} 
\label{tab:teacher_questionnaire_4}
\renewcommand{\arraystretch}{1.15}

\begin{tabular}{p{0.48\textwidth} p{0.48\textwidth}}
\toprule
\textbf{English Version} & \textbf{Chinese Original} \\
\midrule
\multicolumn{2}{l}{\textbf{Dimension 6. Perceived AI Capability}}\\
\midrule

\begin{minipage}[t]{0.48\textwidth}
\textbf{T1.6. Perceived AI Capability}\\
Response scale:\\
(a) Strongly disagree\\
(b) Disagree\\
(c) Neutral\\
(d) Agree\\
(e) Strongly agree\\[4pt]
Items:\\
-- Using generative AI in teaching makes it easier to win awards or receive recognition in teaching competitions.\\
-- I understand that generative AI has information retrieval, analysis, and decision-making capabilities similar to (or even beyond) those of humans.\\
-- I understand that generative AI has multi-turn dialogue, contextual awareness, and self-correction capabilities.\\
-- I understand that generative AI can generate content (text, images, code, audio/video, etc.) at or even beyond human-level quality.
\end{minipage}
&
\begin{CJK}{UTF8}{gbsn}
\begin{minipage}[t]{0.48\textwidth}
\textbf{T1.6. 人工智能能力感知}\\
选项：\\
(a) 非常不同意\\
(b) 不太同意\\
(c) 一般\\
(d) 比较同意\\
(e) 非常同意\\[4pt]
项目：\\
-- 在教学中使用生成式人工智能更容易在各类教学能力比赛中获奖或取得荣誉\\
-- 我了解到生成式人工智能具有类似（甚至超越）“人类”的信息检索、分析与决策能力\\
-- 我了解到生成式人工智能具备多轮对话、上下文感知与自我纠正能力\\
-- 我了解到生成式人工智能可以生成达到（甚至超越）“人类水平”的内容（如图文、代码、音视频等）
\end{minipage}
\end{CJK}
\\
\midrule

\multicolumn{2}{l}{\textbf{Dimension 7. Perceived Supportive Conditions for AI Use}}\\
\midrule

\begin{minipage}[t]{0.48\textwidth}
\textbf{T1.7. Perceived supportive conditions for AI use}\\
Response scale:\\
(a) Strongly disagree\\
(b) Disagree\\
(c) Neutral\\
(d) Agree\\
(e) Strongly agree\\[4pt]
Items:\\
-- The existing equipment and resources at my school (e.g., integrating large models into existing digital teaching systems or deploying local large models) can meet my needs for using generative AI.\\
-- I can conveniently access training on generative AI and find it helpful.\\
-- When I encounter problems while using generative AI, I can conveniently obtain technical or instructional support.
\end{minipage}
&
\begin{CJK}{UTF8}{gbsn}
\begin{minipage}[t]{0.48\textwidth}
\textbf{T1.7. 人工智能使用支持条件感知}\\
选项：\\
(a) 非常不同意\\
(b) 不太同意\\
(c) 一般\\
(d) 比较同意\\
(e) 非常同意\\[4pt]
项目：\\
-- 学校现有的设备与资源（如将大模型接入现有数字教学系统，或本地部署大模型）能够满足我使用生成式人工智能的需求\\
-- 我能够方便地获得生成式人工智能相关培训，并认为这些培训具有实际帮助\\
-- 当我在使用生成式人工智能过程中遇到问题时，能够方便地获得技术或教学支持
\end{minipage}
\end{CJK}
\\

\bottomrule

\end{tabular}
\end{table*}

\begin{table*}
\centering
\caption{Student Questionnaire Items Used in the Study (Bilingual)} 
\label{tab:student_questionnaire_1}
\renewcommand{\arraystretch}{1.15}

\begin{tabular}{p{0.48\textwidth} p{0.48\textwidth}}
\toprule
\textbf{English Version} & \textbf{Chinese Original} \\
\midrule

\multicolumn{2}{l}{\textbf{Dimension I. AI Literacy}} \\
\midrule

\begin{minipage}[t]{0.48\textwidth}
\textbf{S1.1. Which description best defines generative artificial intelligence?}\\
(a) AI that generates new content such as text, images, or music by learning from existing data\\
(b) AI designed to improve search engine retrieval speed and accuracy\\
(c) AI focused on real-time language translation\\
(d) AI mainly used for managing and organizing large databases
\end{minipage}
&
\begin{CJK}{UTF8}{gbsn}
\begin{minipage}[t]{0.48\textwidth}
\textbf{S1.1. 以下哪项对“生成式人工智能”的描述最准确？}\\
(a) 通过学习现有数据创建文本、图像或音乐等新内容的人工智能\\
(b) 一种旨在提高搜索引擎中数据检索速度和准确性的人工智能\\
(c) 一种专注于实时翻译语言的人工智能\\
(d) 主要用于管理和组织大型数据库的人工智能
\end{minipage}
\end{CJK}
\\
\midrule
\begin{minipage}[t]{0.48\textwidth}
\textbf{S1.2. In which task can generative AI typically achieve relatively high accuracy?}\\
(a) Predicting stock market trends\\
(b) Making moral decisions in complex situations\\
(c) Diagnosing rare diseases\\
(d) Generating human-like text based on prompts
\end{minipage}
&
\begin{CJK}{UTF8}{gbsn}
\begin{minipage}[t]{0.48\textwidth}
\textbf{S1.2. 生成式人工智能可以在哪项任务中达到较高的准确率？}\\
(a) 预测股市的走势\\
(b) 在复杂情境中做出道德决策\\
(c) 诊断罕见疾病\\
(d) 根据提示词生成类似人类风格的文本
\end{minipage}
\end{CJK}
\\
\midrule

\begin{minipage}[t]{0.48\textwidth}
\textbf{S1.3. What is a common challenge in prompt-based text generation?}\\
(a) Language models only generate binary outputs\\
(b) Models require large amounts of labeled data during use\\
(c) Writing prompts that accurately capture context and nuance\\
(d) Requiring complex feature engineering
\end{minipage}
&
\begin{CJK}{UTF8}{gbsn}
\begin{minipage}[t]{0.48\textwidth}
\textbf{S1.3. 基于提示词进行文本生成时可能遇到的挑战是？}\\
(a) 语言模型只能生成二元输出\\
(b) 使用阶段需要大量标注数据\\
(c) 撰写能够准确捕捉上下文和细微差别的提示词\\
(d) 需要复杂的特征工程
\end{minipage}
\end{CJK}
\\
\midrule

\begin{minipage}[t]{0.48\textwidth}
\textbf{S1.4. When using generative AI for marketing copywriting, which strategy is least likely to be effective?}\\
(a) Providing information about target audiences\\
(b) Asking AI to include unique selling points\\
(c) Asking AI to use persuasive language techniques\\
(d) Providing a list of competitors’ products
\end{minipage}
&
\begin{CJK}{UTF8}{gbsn}
\begin{minipage}[t]{0.48\textwidth}
\textbf{S1.4. 使用生成式人工智能撰写营销文案时，以下哪种策略最不可能有效？}\\
(a) 向人工智能提供有关目标受众的信息\\
(b) 要求人工智能包含独特卖点和优势\\
(c) 要求人工智能使用具有说服力的语言技巧\\
(d) 向人工智能提供一份竞争对手产品清单
\end{minipage}
\end{CJK}
\\
\midrule

\begin{minipage}[t]{0.48\textwidth}
\textbf{S1.5. A chatbot frequently provides outdated school policy information. What is the best practice to address this issue?}\\
(a) Establish a user feedback mechanism to flag outdated information\\
(b) Regularly update the chatbot training data with latest policies\\
(c) Route complex or policy-related queries to human staff\\
(d) Conduct a comprehensive review of chatbot performance metrics
\end{minipage}
&
\begin{CJK}{UTF8}{gbsn}
\begin{minipage}[t]{0.48\textwidth}
\textbf{S1.5. 客服机器人频繁提供过时的学校政策信息，最佳做法是？}\\
(a) 建立用户反馈机制以标注过时信息\\
(b) 定期更新聊天机器人的训练数据\\
(c) 将复杂或政策相关问题转交人工客服\\
(d) 全面审查聊天机器人性能指标
\end{minipage}
\end{CJK}
\\

\bottomrule

\end{tabular}
\end{table*}

\begin{table*}
\centering
\caption{Student Questionnaire Items Used in the Study (Bilingual)} 
\label{tab:student_questionnaire_2}
\renewcommand{\arraystretch}{1.15}

\begin{tabular}{p{0.48\textwidth} p{0.48\textwidth}}
\toprule
\textbf{English Version} & \textbf{Chinese Original} \\
\midrule

\begin{minipage}[t]{0.48\textwidth}
\textbf{S1.6. When using large language models to collect information for assignments, how should students treat the provided information?}\\
(a) Use it directly without further verification\\
(b) Verify it using other reliable sources\\
(c) Cross-check it with multiple credible references\\
(d) Consider it less reliable than online resources
\end{minipage}
&
\begin{CJK}{UTF8}{gbsn}
\begin{minipage}[t]{0.48\textwidth}
\textbf{S1.6. 学生使用大语言模型完成作业时，应如何对待其提供的信息？}\\
(a) 可直接使用而无需进一步验证\\
(b) 仍需用其他可靠来源验证信息\\
(c) 应与其他可信参考资料进行交叉比对\\
(d) 回答不如网络资源可信
\end{minipage}
\end{CJK}
\\
\midrule

\begin{minipage}[t]{0.48\textwidth}
\textbf{S1.7. When watching a controversial video posted by a public figure, which feature can confirm that it is NOT AI-generated?}\\
(a) The voice sounds similar to the public figure\\
(b) The video appears professionally produced\\
(c) The video quality is high and transitions are smooth\\
(d) None of the above
\end{minipage}
&
\begin{CJK}{UTF8}{gbsn}
\begin{minipage}[t]{0.48\textwidth}
\textbf{S1.7. 以下哪项特征可以确认视频不是由人工智能生成的？}\\
(a) 声音听起来像本人\\
(b) 视频制作精良\\
(c) 视频质量高且过渡流畅\\
(d) 以上都不是
\end{minipage}
\end{CJK}
\\
\midrule

\begin{minipage}[t]{0.48\textwidth}
\textbf{S1.8. A doctor distrusts an accurate AI recommendation because the reasoning process is unclear. What core issue does this illustrate?}\\
(a) Outdated training data\\
(b) Lack of diversity in training datasets\\
(c) Incorrect input treatment guidelines\\
(d) Model opacity (lack of explainability)
\end{minipage}
&
\begin{CJK}{UTF8}{gbsn}
\begin{minipage}[t]{0.48\textwidth}
\textbf{S1.8. 医生因无法理解AI推理过程而不信任结果，这反映了什么核心问题？}\\
(a) 训练数据过时\\
(b) 数据集多样性不足\\
(c) 输入治疗指南错误\\
(d) 模型不透明性
\end{minipage}
\end{CJK}
\\
\midrule

\begin{minipage}[t]{0.48\textwidth}
\textbf{S1.9. What potential copyright issue may arise when journalists use AI-generated images in commercial articles?}\\
(a) Journalists need to review the licensing policies of the AI platform\\
(b) AI-generated images can always be freely used\\
(c) Journalists must always pay standard licensing fees\\
(d) AI-generated images cannot be used commercially
\end{minipage}
&
\begin{CJK}{UTF8}{gbsn}
\begin{minipage}[t]{0.48\textwidth}
\textbf{S1.9. 记者在商业文章中使用AI生成图像可能涉及哪些版权问题？}\\
(a) 需要查看所使用AI平台的许可政策\\
(b) AI生成图像可永久免费使用\\
(c) 必须支付统一授权费用\\
(d) 不能在商业环境中使用
\end{minipage}
\end{CJK}
\\
\midrule

\begin{minipage}[t]{0.48\textwidth}
\textbf{S1.10. Is the statement ``Sharing personal information with online generative AI tools poses almost no privacy risks'' accurate?}\\
(a) Yes, because transmitted data are encrypted\\
(b) Yes, because AI cannot reproduce private information\\
(c) No, because probabilistic outputs may expose sensitive information\\
(d) No, because quantum computing can easily break encrypted data
\end{minipage}
&
\begin{CJK}{UTF8}{gbsn}
\begin{minipage}[t]{0.48\textwidth}
\textbf{S1.10. ``向在线生成式人工智能工具发送个人信息几乎没有隐私风险''是否正确？}\\
(a) 对，因为传输过程中数据已加密\\
(b) 对，因为AI无法输出私密信息\\
(c) 错，因为概率机制可能暴露敏感信息\\
(d) 错，因为量子计算可轻易破解加密数据
\end{minipage}
\end{CJK}
\\

\bottomrule

\end{tabular}
\end{table*}

\end{document}